\documentclass[usenatbib, hyphens]{mnras}

\usepackage{amssymb}
\usepackage{amsmath}
\usepackage{natbib}
\usepackage[pdftex]{graphicx}
\usepackage[normalem]{ulem}
\usepackage{mathtools}
\usepackage{cuted}

\defcitealias{DL07}{DL07}

\title[Orion Bar as a window to the evolution of PAHs]
{Orion Bar as a window to the evolution of PAHs}
\author[M. S. Murga et al.]
{Maria S. Murga$^{1,2}$\thanks{E-mail: murga@inasan.ru},
 Maria S. Kirsanova$^{1}$, Dmitry S. Wiebe$^{1}$, Paul A. Boley$^{3,4}$ \\
$^{1}$Institute of Astronomy, Russian Academy of Sciences, Pyatnitskaya str. 48, Moscow 119017, Russia\\
$^{2}$Faculty of Chemistry, Lomonosov Moscow State University, Universitetsky pr. 13, Moscow 119234, Russia\\
$^{3}$Moscow Institute of Physics and Technology, 141701, 9 Institutskiy per., Dolgoprudny, Moscow Region, Russia\\
$^{4}$Institute of Natural Sciences and Mathematics, Ural Federal University, 19 Mira Str., 620075 Ekaterinburg, Russia}
\date{Accepted today. Received tomorrow; in original form \today}
\pubyear{2021}

\begin{document}
\label{firstpage}
\pagerange{\pageref{firstpage}--\pageref{lastpage}}\maketitle

\begin{abstract}
We investigate the mid-infrared (IR) emission in the Orion Bar photodissociation region, using archival photometric and spectroscopic observations from UKIRT, {\it Spitzer}, ISO, and SOFIA telescopes. Specifically, we consider flux densities of the emission bands at 3.3, 3.4, 3.6, 6.6, 7.7, 11.2~$\mu$m in several locations and a spectrum from 3 to 45~$\mu$m in one location. We study the behaviour of band flux ratios, which are sensitive to external conditions, as revealed by their variations with the distance from an ionizing source. Assuming that the mid-IR emission arises mostly from polycyclic aromatic hydrocarbons (PAHs), and that a weak emission feature at 3.4~$\mu$m is related to PAHs with extra hydrogen atoms (H-PAHs), we trace variations of the ratios using a model for PAH evolution. Namely, we estimate how populations of PAHs of different sizes, hydrogenation and ionization states change across the Orion Bar over a time interval approximately equal to its lifetime. The obtained ensembles of PAHs are further used to calculate the corresponding synthetic spectra and band flux densities. The model satisfactorily describes the main features of the ratios $I_{3.6}/I_{11.2}$, $I_{7.7}/I_{11.2}$,  $I_{7.7}/I_{3.6}$ and $I_{3.3}/I_{3.4}$. We conclude that the best coincidence between modelling and observations is achieved if C loss of PAHs is limited by the number of carbon atoms $N_{\rm C}=60$, and the band at 3.4~$\mu$m may indeed be attributed to H-PAHs. We confirm that large cations dominate at the surface of the PDR but small neutral PAHs and anions are abundant deeper in the molecular cloud.
\end{abstract}


\section{Introduction}

An association of the observed strong mid-infrared (mid-IR) emission in the interstellar medium (ISM) with polycyclic aromatic hydrocarbon molecules (PAHs) had been assumed since early studies by \citet{leger84, 1984ApJ...277..623S, allamandola85}. While some other candidates are also possible \citep[e.~g.][]{kwok11, jones13}, the relation of the mid-IR band emission with PAHs is now widely accepted. Each mid-IR band can be attributed to a specific chemical bond, and certain groups of PAHs are generally responsible for the flux of the particular band according to experimental data and theoretical modelling of PAH spectra made over the last thirty years. In particular, ratios of various band fluxes ($F_{\rm band}$) trace the average size and ionization state of emitting PAHs. Specifically, the band at 3.3~$\mu$m arises mostly from vibrations of the aromatic C--H bond of small neutral PAHs, the bands at 6.2, 7.7, 8.6~$\mu$m are related to vibrations of a C--C bond in ionized PAHs, and the bands at 11.2 and 12.6~$\mu$m arise from the out-of-plane vibrations of a C--H bond in neutral PAHs or anions \citep{allamandola89, bakes01a, DL07, ricca12, croiset16}. The ratio of the band fluxes $F_{3.3}/F_{11.2}$ increases with PAH size \citep{allamandola89, schutte93, croiset16, marag20}, while the ratios $F_{6.2}/F_{11.2}$, $F_{7.7}/F_{3.3}$, $F_{7.7}/F_{11.2}$ correlate with PAH ionization state \citep{bakes01a, boersma18, marag20}. 

Apart from these main bands above, there are also some other bands, which provide additional information about the emitting PAHs. For example, the band at 3.4~$\mu$m is believed to arise from vibrations of the aliphatic C--H bond, which can exist in PAHs with extra hydrogen atoms (super-hydrogenated PAHs or H-PAHs) \citep{bernstein96, sandford13}. In this way, the ratio $F_{3.3}/F_{3.4}$ apparently reflects the ratio between aromatic and aliphatic groups in carbonaceous species, whether they are PAHs with additional aliphatic bonds or particles of some other kind. Aliphatic bonds are less stable than aromatic ones, therefore this ratio may provide key information about PAH evolution, because it might be indicative of specific (and likely unstable) forms of carbonaceous grains surviving under particular conditions. We note that the 3.4~$\mu$m band may also appear due to anharmonicity, but not in any specific PAH state \citep{barker87, maltseva18, chen18a}.

Ratios between the band fluxes vary from one object to another and within one object as outer conditions change. The changes in the PAH ionization state alone cannot explain the sharp variations seen in the mid-IR band fluxes \citep[see e.~g.][]{kassis06}. These variations must be accompanied by changes in the PAH hydrogenation state, size and/or molecular structure.

Today, a general picture of PAH (and PAH-related grains) evolution can be described. In molecular clouds, PAHs can exist either in isolation, or they may stick to each other via weak Van~der~Waals bonds in the form of PAH clusters. Individual PAHs may have small sizes, and also unstable linear geometry. In addition, they can have extra hydrogen atoms, therefore the complicated aromatic-aliphatic complexes may form due to coagulation. Moreover, some heavy elements may adsorb onto PAHs, and these PAHs cease to be pure hydrocarbons. PAHs evolve in the ISM due to the influence of ultraviolet (UV) radiation \citep[see, e.~g.][]{wiebe14, croiset16}. To describe the evolution, several models and approaches have been developed. \citet{allain96_1, visser07, murga16a} consider the photo-destruction of PAHs due to the loss of hydrogen and the acetylene group. \citet{lepage01, montillaud13} focus on the evolution of hydrogenation and charge states. \citet{berne15, parneix17} model more specific processes like isomerisation and the subsequent formation of fullerenes. \citet{montillaud14} study fragmentation of the PAH clusters, while \citet{chen18c} consider the growth of PAHs or their clusterisation. Strong UV flux destroys small and unstable PAHs, along with PAH clusters and complexes \citep[see, e.~g.][]{allain96_1, montillaud13, pavyar13}. Large and stable PAHs survive, but lose their peripheral hydrogen atoms and undergo the process of ring defect formation and isomerisation, including the transition to other allotropic forms \citep[e.~g. fullerenes or nanotubes][]{berne15, chen19_nano}. According to this scenario, only stable large PAHs with a low fraction of hydrogen atoms, as well as fullerenes and nanotubes, should be observed close to sources of the strong UV radiation. At the same time, the existence of small and H-PAHs under these conditions is not expected.

As the life-cycle of PAHs and other small carbonaceous grains strongly depends on the UV~field, a bright and nearby photodissociation region (PDR) with simple geometry becomes an ideal laboratory to test theoretical ideas and models, and also to apply the results of laboratory experiments for the interpretation of observational results. The Orion Bar PDR is a prototypical example of the highly irradiated edge of a molecular cloud, where physical conditions vary dramatically across the cloud -- \ion{H}{II} region interface \citep{tielens85b, Tielens_1993, 2016Natur.537..207G}. Due to its proximity \citep[$\sim414$~pc,][]{menten07} and brightness, an enormous amount of observational data on this object has been accumulated in various spectral ranges, which makes the Orion Bar a versatile test ground for various PDR models. Wide PAH emission bands have been detected and studied in the Orion Bar PDR by, for example, \citet{aitken79, geballe89, schutte93, bregman94, allamandola99, cesarsky00, bakes01a, bakes01b, verstraete01, kassis06, arab12, salgado16}.

The main incentive for this study is to check whether we can reach the consistency between a qualitative scenario of PAH evolution, based on the mid-IR observations of the Orion Bar~PDR, and up-to-date laboratory results and theoretical models on the photo-destruction of PAHs. We consider only the processes related to photo-destruction, but not the formation of PAH clusters and isomerisation, since these processes have not been quantitatively described. We apply a PAH evolution model to the physical and chemical conditions in the Orion Bar PDR, assessed with a chemo-dynamical model, and follow the variations of PAH size, hydrogenation level, and charge throughout the region. Using recent experimental and theoretical PAH IR properties, we calculate synthetic spectra across the PDR, paying particular attention to the 3.4~$\mu$m band, and compare the synthetic spectra with available observational data. 

\section{Observational Data}\label{sect: obs}

In this section we present available observational data on the Orion Bar that have been used in our analysis.

\begin{figure}
	\includegraphics[width=0.45\textwidth]{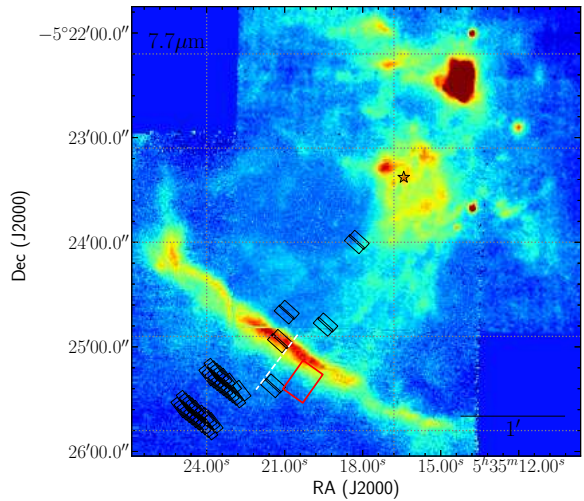}
	\caption{SOFIA FORCAST 7.7-$\mu$m filter image of the Orion bar PDR, with the positions of other mid-IR observations used in this work superimposed. The position of the main ionizing source $\theta^1$~Ori~C is marked by a red star. The positions of the {\it Spitzer} SH-slits (black rectangles), the {\it ISO} slit (red rectangle), the {\it UKIRT} long slit (white dashed line) are shown.} 
	\label{fig: slits}
\end{figure}

\subsection{ISO spectroscopic data}\label{fit ISO}

The ISO spectrum in the 2.4--45.4~$\mu$m range was obtained with the Short Wavelength Spectrometer (SWS) on board of Infrared Space Observatory ({\it ISO}). We downloaded the data, originally presented by \citet{verstraete01}, from the ISO archive\footnote{\url{https://irsa.ipac.caltech.edu/data/SWS}}. The {\it ISO} slit is illustrated by the red rectangle in Fig.~\ref{fig: slits}. The center of the slit corresponds to the H$_2$ emission peak according to \cite{verstraete01}. This position is slightly farther away from the star than the PAH emission peak, seen as the bright bar in the figure. The slit is large enough to cross the Bar itself. The spectrum gives important information about the relationship between the main PAH bands in the Orion Bar, and we use this spectrum further to constrain some PAH characteristics.

The mid-IR emission is characterized by (dust) continuum emission, gaseous lines of hydrogen, argon, neon and other elements, wide PAH emission bands at 3.3, 6.2, 7.7, 8.6, 11.2, 12.6~$\mu$m and some other thinner bands, and some `plateaus', which appear alongside the PAH emission bands, and have a so far unidentified nature \citep{tielens08}. The dust continuum was subtracted by fitting a fifth-order polynomial at wavelengths that do not include any bands or the plateau. Also, we masked bright and narrow gaseous lines. 

The ISO spectrum can be analysed in two ways: 1) by utilising the diagrams suggested by \cite{marag20} to roughly estimate the mean size and charge of PAHs using the ratios $F_{11.2}/F_{3.3}$ and $F_{11.2}/F_{7.7}$; 2) by comparing the observed spectrum with synthetic spectra of PAH mixtures.

In order to use the diagrams from \cite{marag20}, we measured fluxes at 3.3, 7.7, 11.2~$\mu$m in the same manner as they did, i.e. we took the ranges of wavelengths from \cite{marag20}, and integrated the flux densities within these ranges. The obtained ratios of $F_{11.2}/F_{3.3}$ and $F_{11.2}/F_{7.7}$ are approximately 16 and 1, respectively. Further, we used the diagnostic diagrams for different mean photon energies. The radiation field in the Orion Bar can be represented as the mean interstellar radiation field for the ISM in the Solar neighbourhood \citep{mmp83}, scaled by a factor $\chi$, with a mean photon energy of about 6~eV.

In order to find the PAH mixture suitable to the spectrum, we performed a spectral fitting procedure analogous to that of \cite{andrews15} using the following procedure: 
\begin{enumerate}
    \item {\em Selection of PAHs for fitting.} Version 3.0 of the NASA Ames PAH IR database includes vibrational properties for more than 3000 PAHs~\citep{boersma14, bauschlicher18}\footnote{\url{https://www.astrochemistry.org/pahdb/}}. Application of the full database for calculations is time consuming. Therefore, we limited the number of PAHs used for fitting. \cite{andrews15} used around 700 species from version 2.0 and showed that this was enough to describe mid-IR spectra. Namely, the mixture of PAHs in three different PDRs may be limited to $\sim30$ molecules. Therefore, we adopt the PAHs from their lists (in total 42 PAHs) as a basis of our selection. The initial set of \citet{andrews15} included only PAHs that consist of more than 20 carbon atoms and do not contain any other elements besides nitrogen. The smaller PAHs were assumed to be unstable in the ISM, while the PAHs with elements like oxygen, iron, etc. are of low abundance. \cite{andrews15} also did not include PAHs with extra hydrogen atoms. Version 3.0 of the PAH database includes dehydrogenated PAHs (dPAHs) with the same number of carbon atoms and approximately the same topology. To reduce the number of dPAHs in the initial set, we imposed the constraint that no more than two PAHs with a certain number of carbon and hydrogen atoms from one structure family (catacondensed, pericondensed, and irregular)\footnote{We follow the work of \cite{andrews15} in classification.} can be included in the set, independent of their exact geometry. This constraint significantly reduced the selection, and may affect the fitting results, as each PAH has different optical properties (though within one family they are quite similar). We added large pericondensed PAHs with more than 100 carbon atoms in our selection since we found these PAHs are dominant from the diagnostic diagrams (see \ref{sum of obs}). The total number of the selected PAHs is therefore $\sim200$. 
    
    \item {\em Calculation of absorption cross-sections of the sample PAHs.} The list of bands, their central wavelengths, and intensities were taken from the PAH database. We used a Drude profile, following to \citet{DL07} (hereinafter \citetalias{DL07}). As in the work of \cite{andrews15}, both the band FWHM and shift in frequency space (`redshift' in that work) were adopted to be 15~cm$^{-1}$. 
    
    \item {\em Calculation of synthetic spectra of the PAHs.} The UV~field near the location the ISO spectrum $\chi\approx10^4$ based on the FIR emission map presented by \cite{goicoechea15}. We used the method with the multi-photon heating mechanism \citep[][]{pavyar12} to find the temperature distribution for each molecule.
    
    \item {\em Alignment of synthetic and observational spectra.} We reduced the resolution of the spectra to $\lambda/\Delta\lambda=1000$ (approximately the lowest resolution of the ISO spectrum) and performed fitting using the spectra normalised to their maximum values in the range from 5.5 to 15~$\mu$m. The fitting range did not include the bands at 3.3 and 3.4~$\mu$m, as anharmonic effects may be important for this feature \citep[][]{maltseva18, chen18a, chen18b}, but they are not considered in the PAH database. We subtracted the dust continuum from the ISO spectrum, but did not remove the plateau.
    
    \item {\em Fitting procedure.} We used the {\tt leastsq} routine from {\tt Scipy} library for the {\tt Python} language to find the best solution. The results are sensitive to the initial adopted weights of PAHs (or abundances), therefore we performed $\approx 1000$ iterations with random weights from 0 to 1 and selected the iteration with the minimum residual between observed and synthetic spectrum as a result of the fitting. 
\end{enumerate}

\subsection{Spitzer spectroscopic data}\label{Spitzer spectroscopic data}

We use archival spectroscopic data from the {\it Spitzer} Data Archive\footnote{\url{https://sha.ipac.caltech.edu/applications/Spitzer/SHA/}}. The spectra were obtained with the IRS instrument \citep[the Infrared Spectrograph,][]{houck04}. We combine data from different observational programs: ID$=$45 (PI: Thomas Rhoellig),  ID=93, 120 (PI: Dale Cruikshank), ID$=$1094 (PI: Francisca Markwick-Kemper), ID$=$50082 (PI: Robert Rubin). We take only SH (Short-High) slit spectra, as they are positioned perpendicular to the direction from the ionized stars to the molecular cloud, and parallel to the ionization front (IF), therefore we can follow the changes of mid-IR emission as the distance from ionising source increases. The {\it Spitzer} data behind the IF toward to the molecular cloud have been presented by \cite{rubin11, boersma12}. In total, we use 46 slits, which overlap with the SOFIA photometric images (see Fig.~\ref{fig: slits}). The length and the width of each slit are 11.3\arcsec{} and 4.7\arcsec{}, respectively. The spectra cover the wavelength range from 9.89~$\mu$m to 19.51~$\mu$m and have a spectral resolution of $\lambda/\Delta \lambda=600$. We use the post-BCD level product, which does not require any further reduction. As can be seen in Fig.~\ref{fig: slits}, only two slits share the same position at locations closer to the ionizing stars. Therefore, the measurements in these points may be affected by gas and dust heterogeneity of the object as some clumps can fall into these specific slits. The measurements behind the IF were done in several locations, therefore they are quite accurate, thanks to the overlapping slits.

Following \citet{boersma10, boersma12}, we remove the dust continuum and underlying plateau from the spectra to measure the flux of the PAH~11.2~$\mu$m band. In order to find the dust continuum, we fit the points at $\sim$10, 15, and 19.5~$\mu$m by a third-order polynomial function. After subtracting the obtained dust continuum, we fit the points at $\sim$10.6, 10.8, 11.8, 13.4, 13.8, 14.4~$\mu$m by a fifth-order polynomial function to fit the plateau, and also subtract it from the spectrum.

\begin{figure*}
	\includegraphics[width=0.45\textwidth]{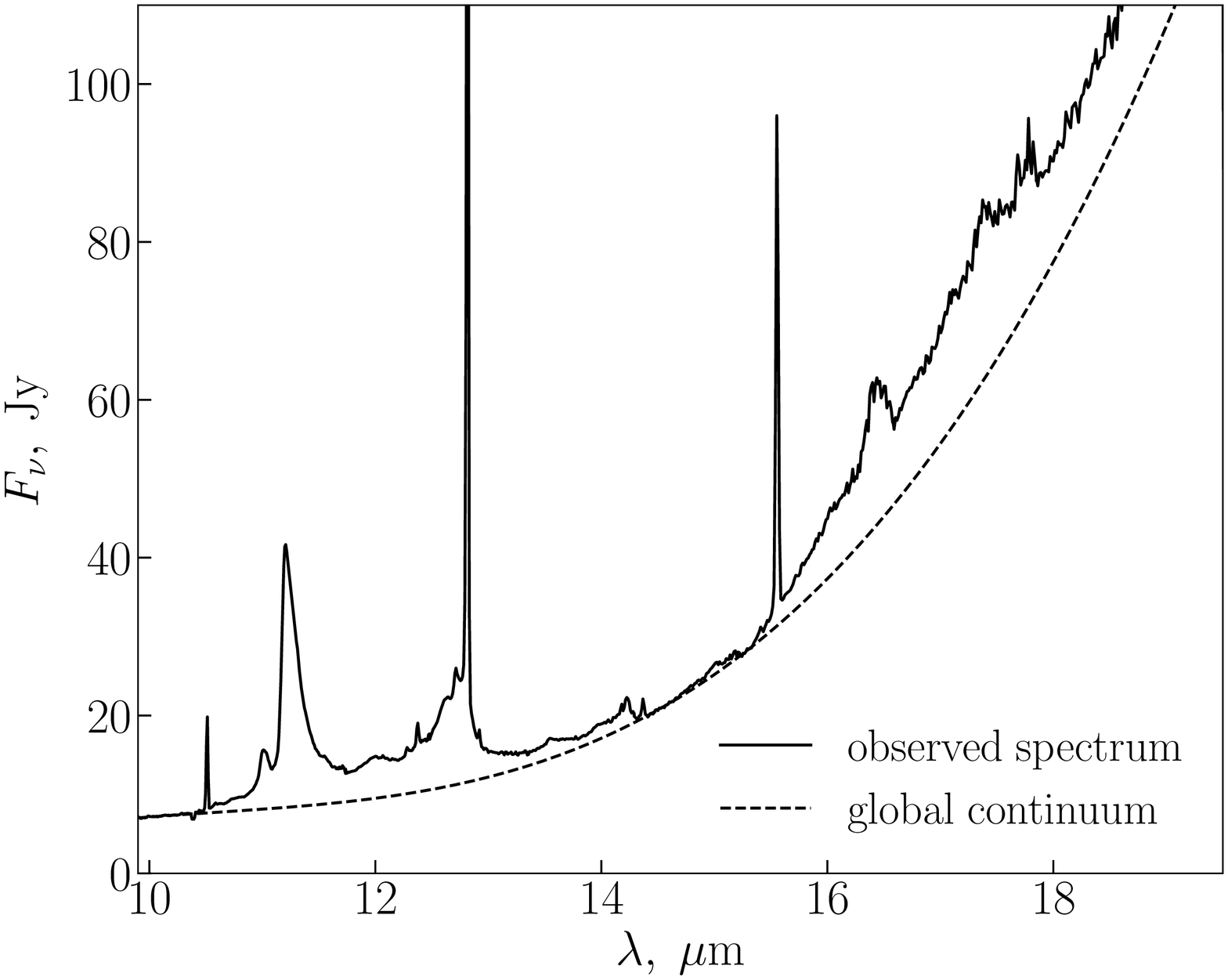}
	\includegraphics[width=0.45\textwidth]{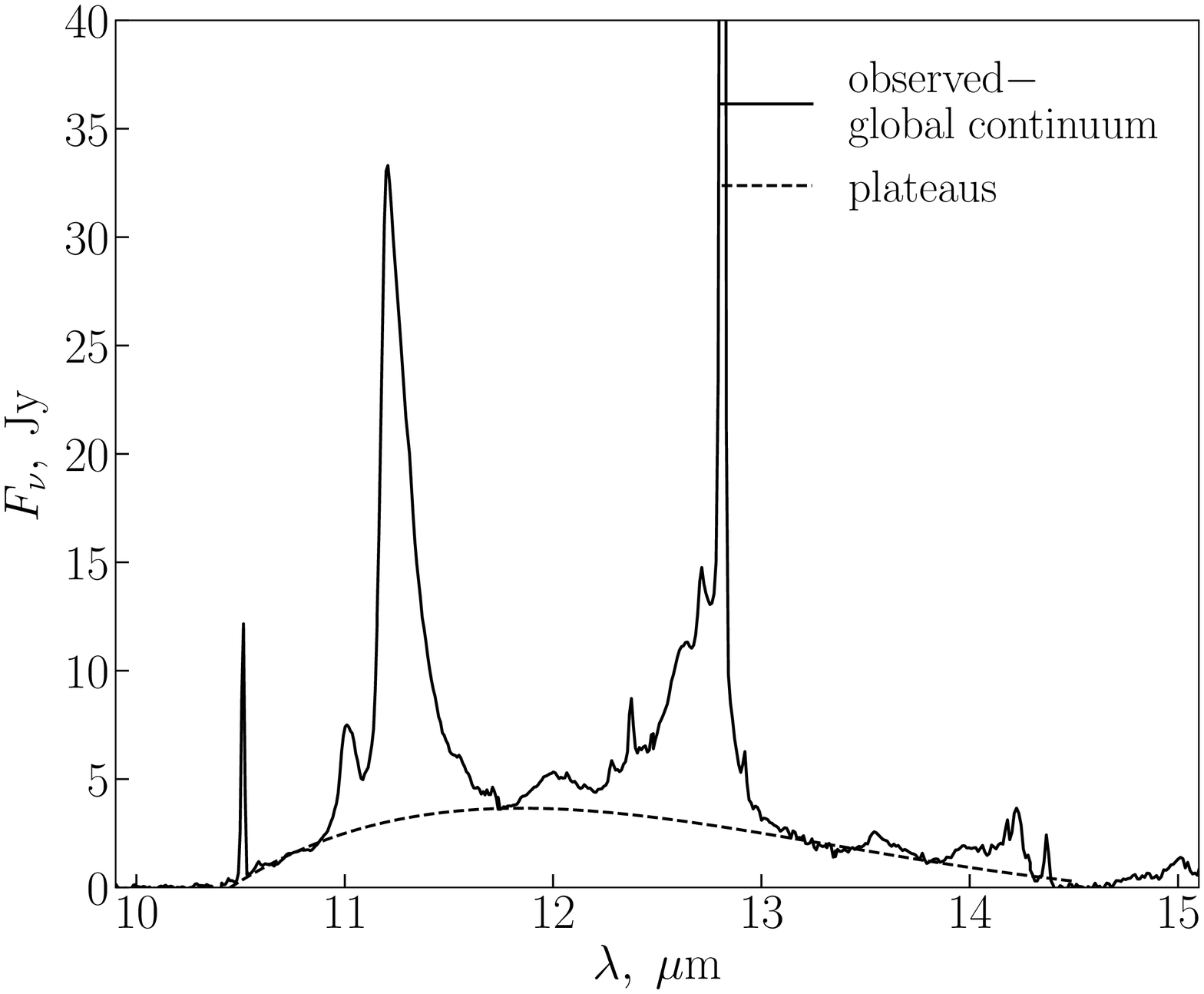}
	\caption{An example of the {\it Spitzer} spectra obtained with the SH-slit spectroscopy. On the left, the global dust continuum is presented by a dashed line. On the right, the PAH plateau underlying the bands at 11--14~$\mu$m is shown by a dashed line.}
	\label{fig: cont}
\end{figure*}

An example of a {\it Spitzer} spectrum is presented in Fig.~\ref{fig: cont}. After continuum and plateau subtraction, we integrate the flux densities over wavelength range which includes the 11.2~$\mu$m band (from 10.9 to 11.6~$\mu$m). Then, we divide the obtained value by the integrated wavelength range to get the average flux density, and by the area of the slit (in units of steradian). Further we average the values obtained in the neighbouring slits at the same distance from the central stars.  Using this method, \cite{boersma12} estimated that the uncertainty of the integrated band flux should be less than 20\%, especially for the bright 11.2~$\mu$m-band. We adopt the uncertainty level of 20\%, noting that averaging over slits may lead to some additional uncertainties.
 
\subsection{UKIRT spectroscopic data}
\label{UKIRT spectroscopic data}

Spectroscopic observations of the Orion Bar PDR in the range from 3.0 to 3.7~$\mu$m were carried out at the United Kingdom Infrared Telescope (UKIRT). The results of these observations were presented by \cite{sloan97}. In total, 32 spectra were obtained along the direction from the ionizing stars toward the molecular cloud with long-slit spectroscopy using a 40\arcsec{} slit. The approximate location of the long slit is shown in Fig.~\ref{fig: slits}. The slit crosses the IF and partially covers the molecular cloud. We chose data obtained on Oct 28, 1995, when the width of the slit was 2.5\arcsec. 

\begin{figure*}
	\includegraphics[width=0.45\textwidth]{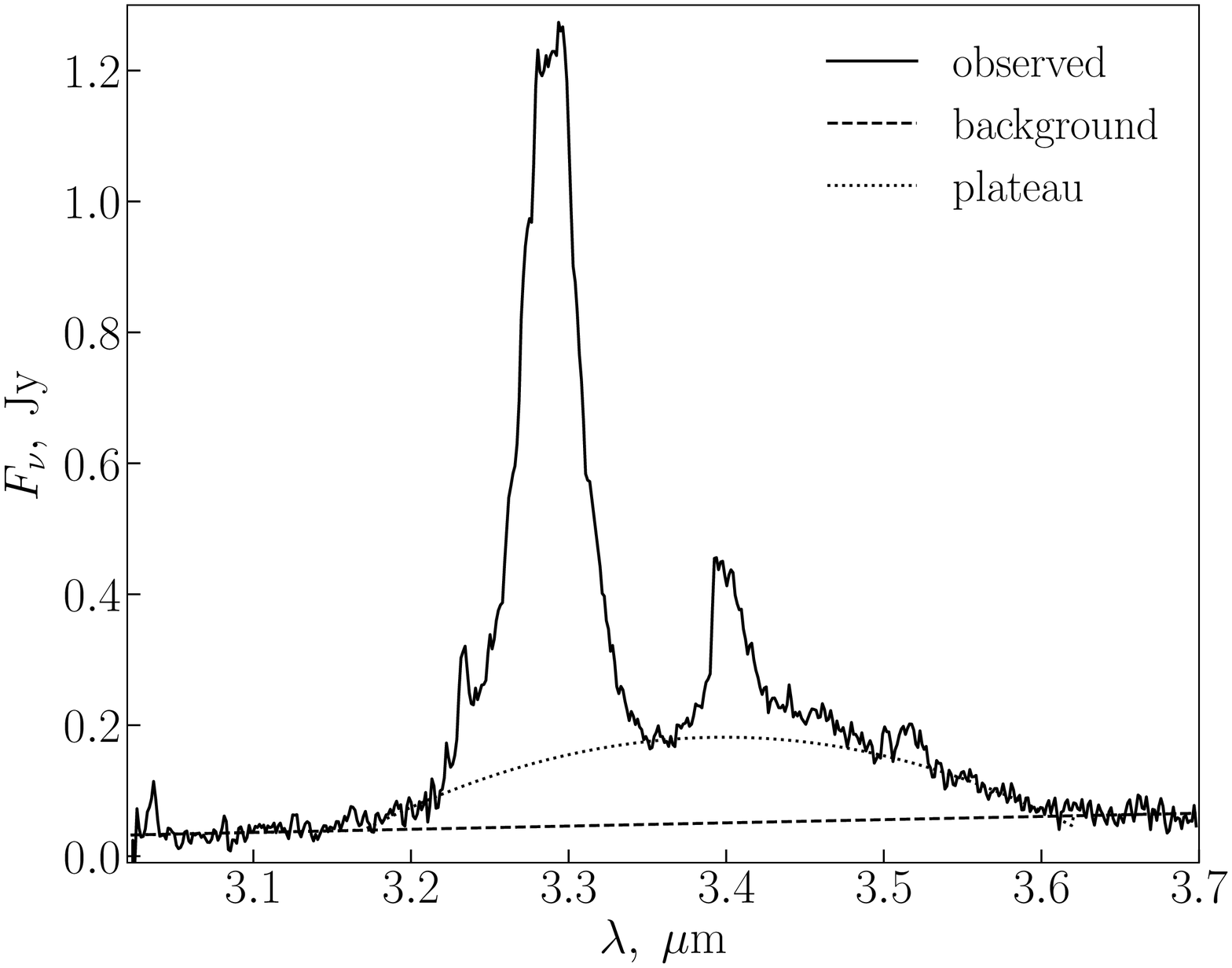}
	\includegraphics[width=0.45\textwidth]{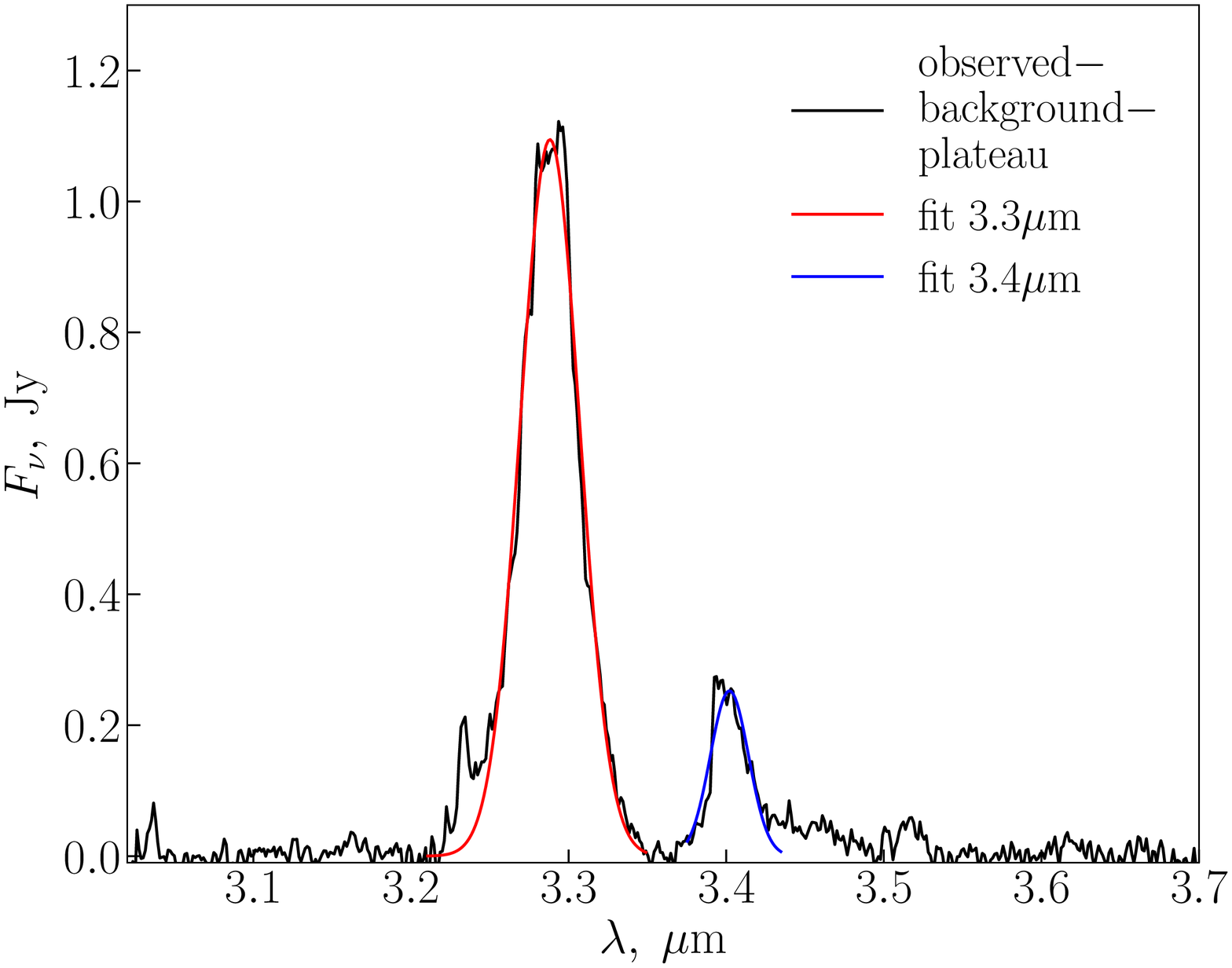}
	\caption{Left: an example of a UKIRT spectrum (the solid line). The continuum fit is shown with a dashed line, and the fit to the plateau is shown with a dotted line. Right: the same spectrum as in the left panel, with the continuum and plateau subtracted. Gaussian fits to the band emission at 3.3 and 3.4~$\mu$m are shown with red and blue lines, respectively.}
	\label{fig: ukirt}
\end{figure*}

To measure flux densities in the 3.3 and 3.4~$\mu$m bands, we estimate the background continuum and the plateau underlying the band emission, subtracted them from the spectra, and fit the bands with a Gaussian function to get the band flux densities. The background continuum was approximated by a linear function in wavelength ranges that do not include any bands ($\lambda<3.15~\mu$m and $\lambda>3.6~\mu$m). The plateau was approximated with a fourth-order polynomial function, with parameters found by fitting over 3 short wavelength segments around $\lambda\approx3.2$, 3.35, 3.55~$\mu$m. An example of a spectrum with the estimated background and the plateau is shown in Fig.~\ref{fig: ukirt} (left). An example of the fitting of the bands by a Gaussian function is shown in Fig.~\ref{fig: ukirt} (right). We integrate over the wavelength of the bands and then divide by the corresponding length of wavelengths to get average flux densities. The uncertainties of the band flux densities were estimated using a bootstrap method. We do not show the flux densities and uncertainties, as they are very close to the results of \cite{sloan97}. 

\subsection{SOFIA and Spitzer photometric data}

We use photometric data obtained with the SOFIA FORCAST\footnote{{\tt\url{https://www.sofia.usra.edu/science/instruments/forcast}}} \citep[]{herter18} instrument at 6.6 and 7.7~$\mu$m,
and the photometric image at 3.6~$\mu$m from the {\it Spitzer} telescope archive (programme ID $=50$), obtained with the IRAC instrument\footnote{{\tt\url{https://irsa.ipac.caltech.edu/data/SPITZER/docs/irac}}}\citep{fazio04}.

We performed photometry in two ways. In the first case, we took the {\it Spitzer} slits, summed the fluxes in all the pixels which fell inside each slit, and divided these sums by the area of the slit obtaining average values over the slit areas. The Spitzer and SOFIA photometric data provide average flux densities in each pixel. In the second case, we measured flux densities in stripes with a width of 1.2~arcsec, parallel to the IF of the Orion Bar and perpendicular to the direction from the star towards the molecular cloud. We chose the length of the stripes so that they are comparable to the length of the brightest part of the Bar. The optimal length was found to be $54\arcsec$. 

The filter centred at 7.7~$\mu$m corresponds closely to a strong PAH band. We estimated the contributions of the dust continuum and plateau at this band, using the flux density of the 6.6~$\mu$m filter as the background value, following the work of \cite{salgado12}. The standard deviation of the pixel flux densities inside the slits are adopted as the uncertainties of the flux density measurements at 3.6 and 7.7~$\mu$m. To compare measurements at 3.6 and 7.7~$\mu$m to each other, we convolved the 7.7~$\mu$m image to the resolution of the 3.6~$\mu$m image, which has a lower resolution. The measurements at 7.7~$\mu$m obtained from the convolved image are designated as $F_{7.7}^{\rm conv}$. The results of the measurements at 3.6 and 7.7~$\mu$m ($F_{3.6}$ and $F_{7.7}$) in the slits are presented in Table~\ref{tab: fluxes}, along with $F_{11.2}$, obtained from the {\it Spitzer} spectroscopy (see above).

\begin{table}
\caption{Distances of the {\it Spitzer} SH-slits from the ionizing source and the average flux densities at wavelengths 3.6, 7.7, and 11.2~$\mu$m measured in the slits.}
\label{tab: fluxes}
\begin{center}
\begin{tabular}{cccc}
\hline
Distance, & $F_{3.6}$  &$F_{7.7}$ &$F_{11.2}$  \\
arcsec  & \multicolumn{3}{c}{$10^8\cdot$Jy~sr$^{-1}$} \\ \hline
 44.1 &   1.3 $\pm$  0.2 &  30.6 $\pm$   0.5  &  18.2 $\pm$   3.7  \\
 47.7 &   1.1 $\pm$  0.1 &  27.1 $\pm$   0.4  &  17.0 $\pm$   3.4  \\
 94.1 &   1.3 $\pm$  0.1 &  30.9 $\pm$   0.5  &  26.9 $\pm$   5.4  \\
 97.4 &   1.3 $\pm$  0.1 &  32.3 $\pm$   0.5  &  28.9 $\pm$   5.8  \\
100.7 &   1.0 $\pm$  0.1 &  19.1 $\pm$   0.3  &  11.6 $\pm$   2.3  \\
104.6 &   0.3 $\pm$  0.1 &  18.7 $\pm$   0.4  &  14.3 $\pm$   2.8  \\
116.2 &   3.0 $\pm$  0.1 &  114.5 $\pm$   1.3  &  68.8 $\pm$  13.2  \\
119.8 &   1.4 $\pm$  0.1 &  114.6 $\pm$   1.3  &  69.2 $\pm$  13.8  \\
139.5 &   1.9 $\pm$  0.1 &  24.1 $\pm$   0.4  &  22.1 $\pm$   4.4  \\
143.2 &   1.7 $\pm$  0.1 &  22.5 $\pm$   0.4  &  19.9 $\pm$   4.0  \\
155.5 &   0.6 $\pm$  0.1 &  16.1 $\pm$   0.3  &  14.6 $\pm$   3.0  \\
161.6 &   1.1 $\pm$  0.1 &  12.0 $\pm$   0.3  &  12.9 $\pm$   2.6  \\
 
\hline
\end{tabular}
\end{center}
\end{table}

\subsection{Summary of observational data}
\label{sum of obs}
\begin{figure}
	\includegraphics[width=0.45\textwidth]{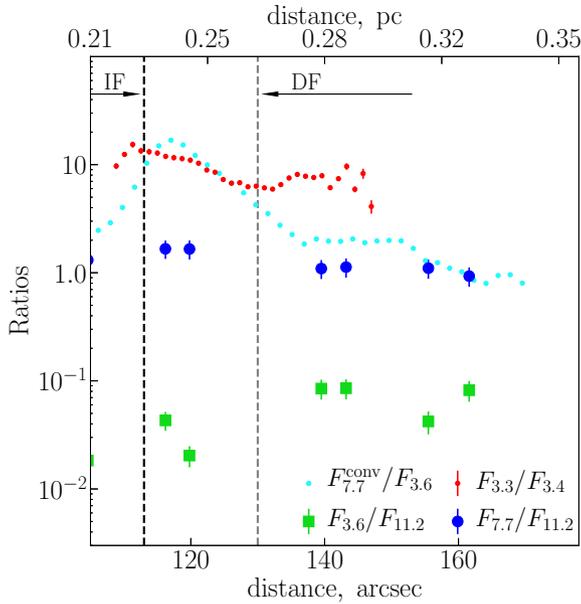}
	\caption{The observed ratios between band flux densities. The ratios $F_{3.6}/F_{11.2}$, $F_{7.7}/F_{11.2}$, $F_{3.3}/F_{3.4}$ and $F_{7.7}^{\rm conv}/F_{3.6}$ are marked by various symbols (green squares, blue circles, red circles and light blue circles, respectively). The ionization and dissociation fronts are marked by black and gray dashed lines, respectively.}
	\label{fig: sum_obs}
\end{figure}

We combine all the described IR flux densities as a function of the distance to the ionizing stars in Fig.~\ref{fig: sum_obs}. The IF is located at 113\arcsec{} from $\theta^1$~Ori~C \citep[][]{O_Dell_2000}. The distance between the H$_2$ dissociation front (DF) and the IF is 17\arcsec{} \citep[][]{1996A&A...313..633V}. 

The ratios have their extrema near the IF. The ratios $F_{7.7}^{\rm conv}/F_{3.6}$ and $F_{7.7}/F_{11.2}$ have maximums at the IF, then they decrease towards the molecular cloud, and the trend of decreasing continues in the cloud, while the slope is not as steep as between the IF and DF. The ratio $F_{3.6}/F_{11.2}$ has a minimum right behind the IF and increases towards the molecular cloud. The ratio $F_{3.4}/F_{3.3}$ has a maximum near the IF and then drops to its minimum near the DF. In the molecular cloud this ratio starts growing slowly, although some scatter in the values of the ratios is present.

Using diagrams from \citet{marag20}, we find that PAHs in the Orion Bar contain 120--130 carbon atoms, and that the fractions of cations and neutral PAHs should be approximately equal. These diagrams were based on the absorption cross-sections of PAHs with the appropriate number of hydrogen atoms, i.e. with all peripheral carbon atoms which have a free valence electron connected a hydrogen atom. Therefore, the diagrams may not be suitable for the cases when PAHs are super-hydrogenated or dehydrogenated, as their absorption cross-sections may depend on the hydrogenation level. Therefore, we consider these results as a qualitative constraint.

\begin{figure}
	\includegraphics[width=0.45\textwidth]{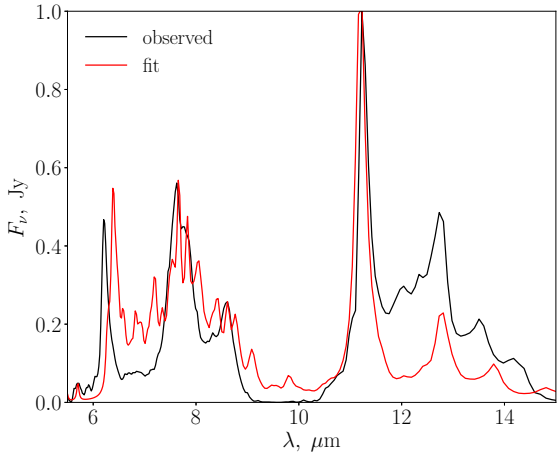}
	\caption{The ISO spectrum of the Orion Bar and the synthetic spectrum of the PAH mixture obtained with the fitting routine.}
	\label{fig: fit}
\end{figure}

We show the results of the ISO spectrum fitting in Fig.~\ref{fig: fit}. The obtained PAH mixture, unique identifier numbers (UID) from the PAH database, normalised contribution weights, and structure of each component are presented in Table~\ref{tab: fit}. The weights indicate the fraction of specific PAHs in the mixture. From Table~\ref{tab: fit}, we conclude that PAHs, producing the ISO spectrum, should contain at least 50--100 C atoms. The major contribution (more than 80\%) comes from large PAHs (C$_{96}$H$_{26}$, C$_{98}$H$_{28}^{+}$, C$_{142}$H$_{30}$), which are responsible for almost all of the main bands. These PAHs can be neutral, positively or negatively charged. There is a fraction (about 17\%) of small catacondensed dehydrogenated PAHs in the mixture, which are required for describing the band at 5.7~$\mu$m, although they also contribute to the 6--8~$\mu$m region. We conclude that the PAH mixture for the Orion Bar PDR can be found but with some residuals especially in the regions 6.5--7, 9--10, 11.7--14~$\mu$m. Obviously, some hydrogenated PAHs traced by the feature at 3.4~$\mu$m should be in the list as well, but we did not consider the range that covers this feature.

\begin{table}
\caption{The PAH mixture obtained with the fitting routine. The table includes a formula of each PAH, its UID, the normalised contribution weight and structure.}
\label{tab: fit}
\begin{center}
\begin{tabular}{cccc}
\hline
Formula & UID & Weight & Structure  \\ \hline
C$_{21}$    & 2372  & 0.048 & Catacondensed  \\
C$_{34}$H$_{20}^+$   & 878  & 0.122 & Catacondensed \\
C$_{80}$H$_{20}$    & 719  & 0.062 & Pericondensed \\
C$_{96}$H$_{28}^-$   & 192  & 0.082 & Pericondensed \\
C$_{96}$H$_{25}$   & 690  & 0.020 & Pericondensed \\
C$_{96}$H$_{26}$   & 692  & 0.273 & Pericondensed \\
C$_{98}$H$_{28}^{+}$   & 568  & 0.243 & Pericondensed \\
C$_{112}$H$_{26}$   & 165  & 0.001 & Pericondensed \\
C$_{142}$H$_{30}$   & 766  & 0.123 & Pericondensed  \\
C$_{146}$H$_{30}$   & 752  & 0.024 & Pericondensed  \\
\hline
\end{tabular}
\end{center}
\end{table}

\section{Model of PAH evolution}\label{sect: model}

\subsection{Main equations of the model}

We model evolution of PAHs using {\tt Shiva} dust destruction model \citep{murga19}, with additional processes and modifications described by \citet{murga20}. We present the main features of the model, focusing on the new points adopted in this work. We consider: the loss of atomic hydrogen, molecular hydrogen, acetylene~(C$_2$H$_{2}$) or diatomic carbon~(C$_2$), as well as addition of hydrogen and carbon atoms. Due to these processes PAH can change their size (and mass) and hydrogenation level.

The hydrogenation level in each bin is characterized by the mass of hydrogen atoms. The minimum value is 0 when PAHs do not contain hydrogen atoms, and the maximum value corresponds to a hypothetical super-hydrogenated state, when PAHs are fully hydrogenated, i.e. all peripheral carbons atoms are bonded with two hydrogen atoms and inner carbon atoms are also bonded with one hydrogen atom. We will also use the ratio between the number of carbon and hydrogen atoms in a PAH molecule, $X_{\rm H}$, as characteristic of the PAH hydrogenation level. We assume a normal hydrogenated state if a PAH possesses the same ratio as in its original state ($X_{\rm H}=X_{\rm H}^{0}$), i.e. each peripheral carbon atom with free valence electron is connected to one hydrogen atom. If $X_{\rm H} > X_{\rm H}^{0}$, the state is considered as super-hydrogenated. If $X_{\rm H} < X_{\rm H}^{0}$, then the state is dehydrogenated. 

The considered PAH mass range is divided into $N_{\rm m}$ bins with borders designated as $m_{\rm b}^{i}$. The mass within each bin includes both carbon and hydrogen atoms. The mass range of hydrogen atoms is divided into $N_{\rm mH}$ bins with borders $m_{\rm Hb}^{ij}$. The number density in each bin is designated as $N_{ij}$ and is calculated from the mass distribution as $dn/dm(m_{\rm b}^{i+1}-m_{\rm b}^{i})$. We solve the following kinetic equation for each $N_{ij}$  
\begin{eqnarray}
\frac{dN_{ij}}{dt} &=& A_{ij+1}^{(1)}N_{ij+1} - A_{ij}^{(1)}N_{ij} \text{ }| \text{ H loss} \nonumber\\ 
                    & + &A_{ij-1}^{(2)}N_{ij-1} - A_{ij}^{(2)}N_{ij} \text{ }| \text{ H addition} \\
                    & + &B_{i+1j}^{(1)}N_{i+1j} - B_{ij}^{(1)}N_{ij} \text{ }| \text{ C loss} \nonumber \\ \nonumber
                    & + & B_{i-1j}^{(2)}N_{i-1j} - B_{ij}^{(2)}N_{ij} \text{ }| \text{ C addition} \nonumber \\ \nonumber 
\label{main_eq}
\end{eqnarray}
where $A^{(1,2)}_{ij}$  and $B^{(1,2)}_{ij}$ are rate coefficients for the processes of hydrogen and carbon mass change, respectively.

The rate coefficients for processes of H loss and addition are calculated as
\begin{equation}
A^{(1,2)}_{ij} = \frac{\varepsilon^{(1,2)}_{ij}}{m_{{\rm Hb}}^{ij+1}-m_{{\rm Hb}}^{ij}}.
\label{Acoef}
\end{equation}
We adopt the following boundary values: $A^{(1)}_{ij}=0$ at $j=1$, and  $A^{(2)}_{ij}=0$ at $j=N_{\rm mH}$, i.e. fully dehydrogenated grains no longer can lose their hydrogen atoms, and fully hydrogenated grains cannot acquire hydrogen atoms. The values $\varepsilon^{(1)}_{ij}$ and $\varepsilon^{(2)}_{ij}$ represent the rate of change of H atoms mass and are expressed as
\begin{eqnarray}
\varepsilon^{(1)}_{ij} &=& \frac{\mu_{\rm H}}{N_{\rm A}}R_{\rm H}^{ij}  \nonumber \\
\varepsilon^{(2)}_{ij}&=& \frac{\mu_{\rm H}}{N_{\rm A}}H^{ij},
\end{eqnarray}
where $R_{\rm H}^{ij}$ and $H^{ij}$ are rates of detachment and attachment of hydrogen atoms, respectively. They are given by \citet{murga20}. We note here that $R_{\rm H}^{ij}$ is the net rate of detachment of all hydrogen atoms which might be detached from a PAH both separately and also inside hydrogen or acetylene molecules. The $\mu_{\rm H}$ value is the molar hydrogen atom mass, and $N_{\rm A}$ is the Avogadro constant.

The rate coefficients for carbon loss and addition are calculated analogously:
\begin{equation}
B^{(1,2)}_{ij} = \frac{\mu^{(1,2)}_{ij}}{m_{\rm b}^{i+1}-m_{\rm b}^{i}}.
\end{equation}

The boundary values of $B_{ij}^{(1)}$ and $B_{ij}^{(2)}$ for $i = N_{\rm m}$ are set to zero: the grains cannot get into the $N_{\rm m}$th bin from any bin with $i>N_{\rm m}$, and they cannot move from this bin to the bins with $i>N_{\rm m}$. $B_{ij}^{(2)}$ is set to zero if $i<1$, i.e., we do not allow new PAHs to appear from scratch, but we allow grains to leave the $i=1$ bin. The values of $\mu_{ij}$ represent the rates of change of C atoms mass and are expressed similar to $\varepsilon^{(1)}_{ij}$ as
\begin{eqnarray}
\mu^{(1)}_{ij} &=& \frac{\mu_{\rm C}}{N_{\rm A}}R_{\rm C}^{ij}  \nonumber \\
\mu^{(2)}_{ij}&=& \frac{\mu_{\rm C}}{N_{\rm A}}C^{ij},
\end{eqnarray}
where $R_{\rm C}^{ij}$ and $C^{ij}$ are the rates of detachment and attachment of carbon atoms, respectively. These values are given by \citet{murga20}. The $\mu_{\rm C}$ is the molar carbon atom mass. 

The rates of C and H loss in our calculations depend on the bond activation energy $E_0$ and the change of entropy due to the bond dissociation $\Delta S$. These parameters are critically important for the modelling. They vary from one PAH to another. Furthermore, they have been determined only for a small set of PAHs. We collect the most relevant values of $E_0$ and $\Delta S$ from the literature. For the H and H$_2$ loss we rely on the parameters adopted by \cite{andrews16} as in the work of \citet{murga20}. The parameters for the C loss are more difficult to choose. The values from \cite{mic10} can be adopted for the normal and dehydrogenated states, as they are supported by experiments, but the parameters for the super-hydrogenated states are less certain. Several experiments have shown the instability of H-PAHs. The more extra H-atoms the H-PAHs have, the less stable they become \citep{wolf16, quitian18, rapacioli18}. We estimate these parameters from the experiment of \cite{rapacioli18}, who investigated fragmentation of a coronene molecule saturated by hydrogen. They measured the yields of fragments for different hydrogenation states and found that the yield of various carbonaceous fragments (C$_x$H$_y$) increases with the number of extra hydrogen atoms, while the survival yield decreases. When the PAHs are fully hydrogenated, the yield of C$_2$H$_2$ is higher by a factor of $\sim50$ relative to the yield in the normal (initial) state. At the same time the survival yield decreases by a factor of 2. 

We can calculate the yields using the expressions from the {\tt Shiva} model \citep{murga19, murga20} and try to find consistency with the experiment of \cite{rapacioli18}. We assume that the parameter $E_0$ changes depending on the hydrogenation level, i.e. we consider $E_0$ as a parameter describing the yield of destruction rather than the activation energy. We adopt that $E_0$ changes with hydrogenation level as a linear function
\begin{equation}
E_0 = a^{\ast} \frac{X_{\rm H}}{X_{\rm H}^{0}}+b^{\ast},
\end{equation}
where $a^{\ast}$ and $b^{\ast}$ are some parameters. To estimate $a^{\ast}$ and $b^{\ast}$ we set the boundary conditions for H and C$_2$H$_2$ loss for hydrogenation states with $X_{\rm H}=X_{\rm H}^{0}$ and $X_{\rm H}=2\cdot X_{\rm H}^{0}$. As for H loss in a PAH in the hydrogenated state with $X_{\rm H}=2\cdot X_{\rm H}^{0}$, we use the mean values for solo and duo hydrogen atoms given by \cite{andrews16}: we adopt $E_0=1.8$, 1.6 and 1.9~eV for anions, neutral and cations, respectively. For C$_2$H$_2$ loss we found the most suitable value for each PAH in each ionization state, such that the yields of carbonaceous fragments and survival are consistent with the measurements of \cite{rapacioli18}. We found 1.9, 1.7 and 2.0~eV for anions, neutral and cations, respectively, in the hydrogenated state with $X_{\rm H}=2\cdot X_{\rm H}^{0}$; i.e. they are higher than the corresponding values for the H loss by 0.1~eV.  The linear functions obtained, $E_0$ and $\Delta S$ for all states are presented in Table~\ref{diss_par}.

\begin{table}
\caption{Dissociation parameters $E_0$~[eV] and $\Delta S$~[cal~K$^{-1}$~mol$^{-1}$].}
\label{diss_par}
\begin{center}
\begin{tabular}{|l|c|c|c|c|}
\hline
Fragment & \multicolumn{2}{c|}{$X_{\rm H}\leq X_{\rm H}^{0}$} & \multicolumn{2}{c|}{$X_{\rm H}> X_{\rm H}^{0}$}      \\
\hline 
                  &    $E_0$   & $\Delta S$     &   $E_0$                                             &   $\Delta S$ \\
\hline
H  ($Z<0$)        &     4.3    &       11.8     &     $-2.5 \frac{X_{\rm H}}{X_{\rm H}^{0}}+6.8$      &   13.3       \\
H  ($Z>0$)        &     4.3    &        11.8    &     $-2.4 \frac{X_{\rm H}}{X_{\rm H}^{0}}+6.7$      &   13.3       \\
H  ($Z=0$)        &     4.3    &       11.8     &     $-2.7 \frac{X_{\rm H}}{X_{\rm H}^{0}}+7.0$      &   13.3       \\
H$_2$             &     3.52   &     --12.69    &        --                                           &    --        \\
C$_2$H$_2$ ($Z<0$)&     4.6    &       10.0     &      $-2.7 \frac{X_{\rm H}}{X_{\rm H}^{0}}+7.3$     &    10.0      \\
C$_2$H$_2$ ($Z>0$)&     4.6    &       10.0     &      $-2.6 \frac{X_{\rm H}}{X_{\rm H}^{0}}+7.2$     &    10.0      \\
C$_2$H$_2$ ($Z=0$)&     4.6    &       10.0     &      $-2.9 \frac{X_{\rm H}}{X_{\rm H}^{0}}+7.5$     &    10.0      \\
\hline
\end{tabular}
\end{center}
\end{table}

\citet{zhen15, zhen16} showed that the interaction of large PAHs with UV photons leads to ionization, rather than fragmentation. Some studies indicate that dehydrogenation most likely occurs with large PAHs containing 70--80 carbon atoms \citep{berne12, zhen14}. As experiments with destruction of large PAHs are lacking, we consider four scenarios of carbon skeleton destruction: a)~PAHs cannot lose their hydrogen and carbon atoms; b)~PAHs can lose hydrogen atoms but cannot lose carbon atoms; c)~PAHs can lose their hydrogen and carbon atoms no matter how large they are, and d)~all PAHs can lose their hydrogen atoms, but only PAHs with $N_{\rm C}<60$ can lose their carbon atoms. 
The cases are schematically described in Table~\ref{tab: cases}. The `a' model represents the initial state of the PAH ensemble. The `b' and `c' models represent extreme cases of carbon skeleton destruction. The `d' model is chosen based on experiments and theoretical investigations revealing the inefficient destruction of the carbon skeleton beyond a critical PAH size ($N_{\rm C}\approx60$ or $a\approx5$~\AA{})\citep{zhen15}. While these experiments were conducted for only the normally hydrogenated PAHs, we adopt this critical size for all PAH states, since no experiments are available for large super-hydrogenated PAH analogues, i.e., we assume that the behaviour of hydrogenated and super-hydrogenated PAHs is similar. If, on the other hand, we assume that the stability properties of super-hydrogenated PAHs are close to properties of hydrogenated amorphous carbon (HAC)\citep[][]{alata14, alata15, duley15}, the assumption on the limiting size is not valid, and the `c' model is more appropriate for the these states.

\begin{table}
\caption{Model variants considered in the work.}
\label{tab: cases}
\begin{center}
\begin{tabular}{|l|c|c|}
\hline
Case & H loss  & C loss\\
\hline
`a' & $-$ & $-$ \\
`b' & $+$ & $-$ \\
`c' & $+$ & $+$ \\
`d' & $+$ & $+$ for $N_{\rm C}<60$ \\
\hline
\end{tabular}
\end{center}
\end{table}

The evolutionary processes of ionized and neutral PAHs proceed with different rates \citep[e.g.][]{allain96_2}. The ionization state is also crucial for computing synthetic spectra, therefore, we calculate the evolution of PAHs depending on their charge, and estimate the PAH charge probability function $f_i(Z)$ for each size bin based on the work of \cite{wd01_charge}.

PAH evolution in UV radiated environments has been modelled previously several times. Some models focus primarily on carbon skeleton destruction \citep{allain96_1}, while others consider only variations of hydrogenation level and charge \citep{lepage01, montillaud13}. Our model, on the other hand, provides a combined approach for all relevant processes for PAHs. Unlike the previous models, we follow the evolution of the full ensemble of PAHs, taking into account that PAHs can move from one group (in size, hydrogenation level) to another.

\subsection{Input parameters for PDR modelling and additional tools}

For the modelling of the PAH evolution, we adopt physical parameters of the Orion Bar PDR from the chemo-dynamical model {\tt MARION} \citep{marion}. The Orion Bar has been modelled many times \citep[e.g.][]{Tielens_1993, 2009ApJ...693..285P, 2009ApJ...701..677S, 2011MNRAS.416.1546A}, but the {\tt MARION} model has several advantages over other models. The dynamics and chemistry are coupled in this model, therefore the model can be used to study how the chemo-dynamical structure of the \ion{H}{II} region, PDR, and molecular gas changes with time \citep[see, e. g.][]{2020MNRAS.497.2651K}. Also, the model takes into account the relative motions of gas and dust \citep[][]{2015MNRAS.449..440A, 2017MNRAS.469..630A} and allows multiple grain populations. We use the modification of the model presented by \cite{kirsanova19}, who used it to describe the H$_2$ and CO dissociation fronts in the Orion Bar.

\begin{figure*}
\includegraphics[width=0.49\textwidth]{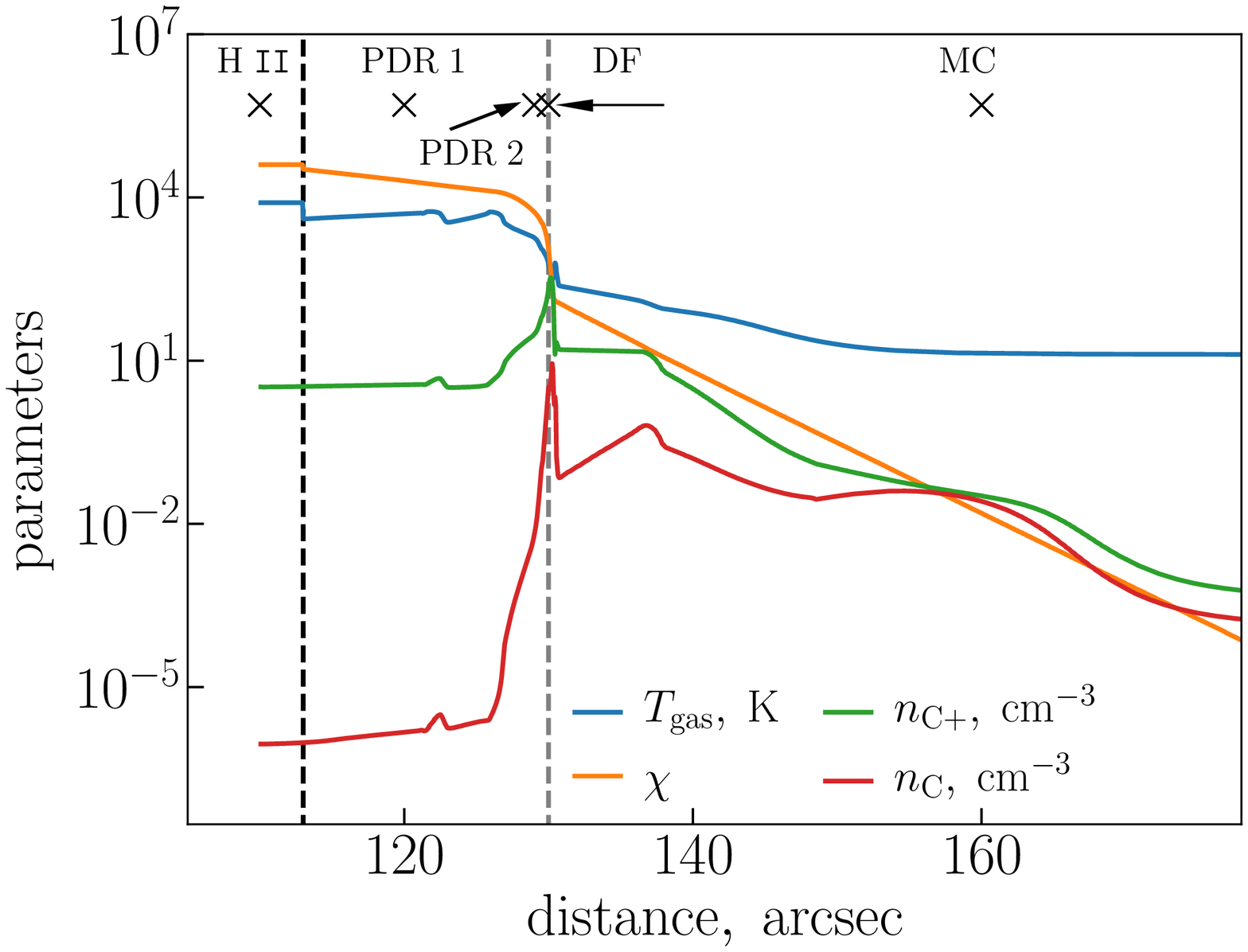}
\includegraphics[width=0.49\textwidth]{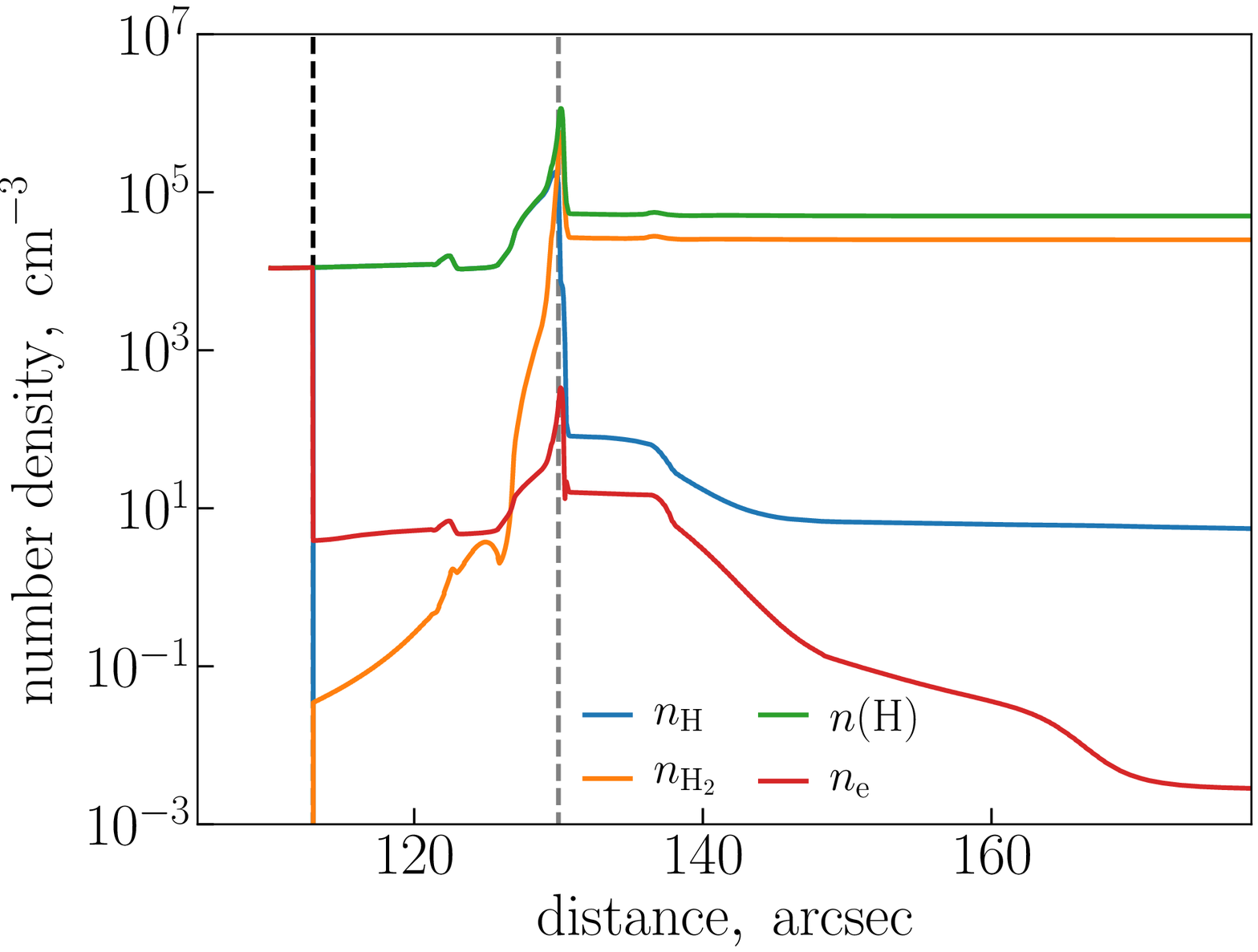}\\
\includegraphics[width=0.49\textwidth]{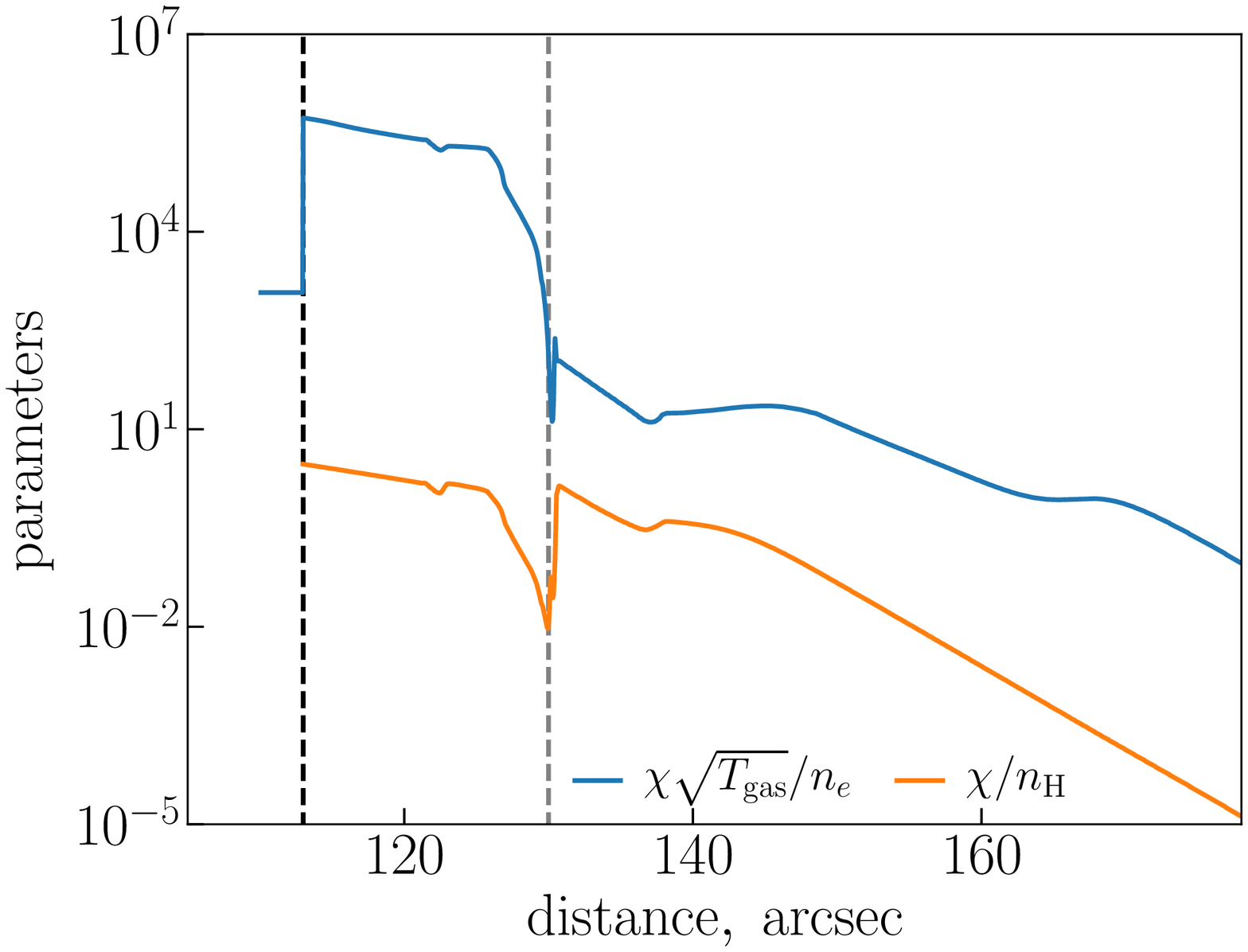}
\includegraphics[width=0.49\textwidth]{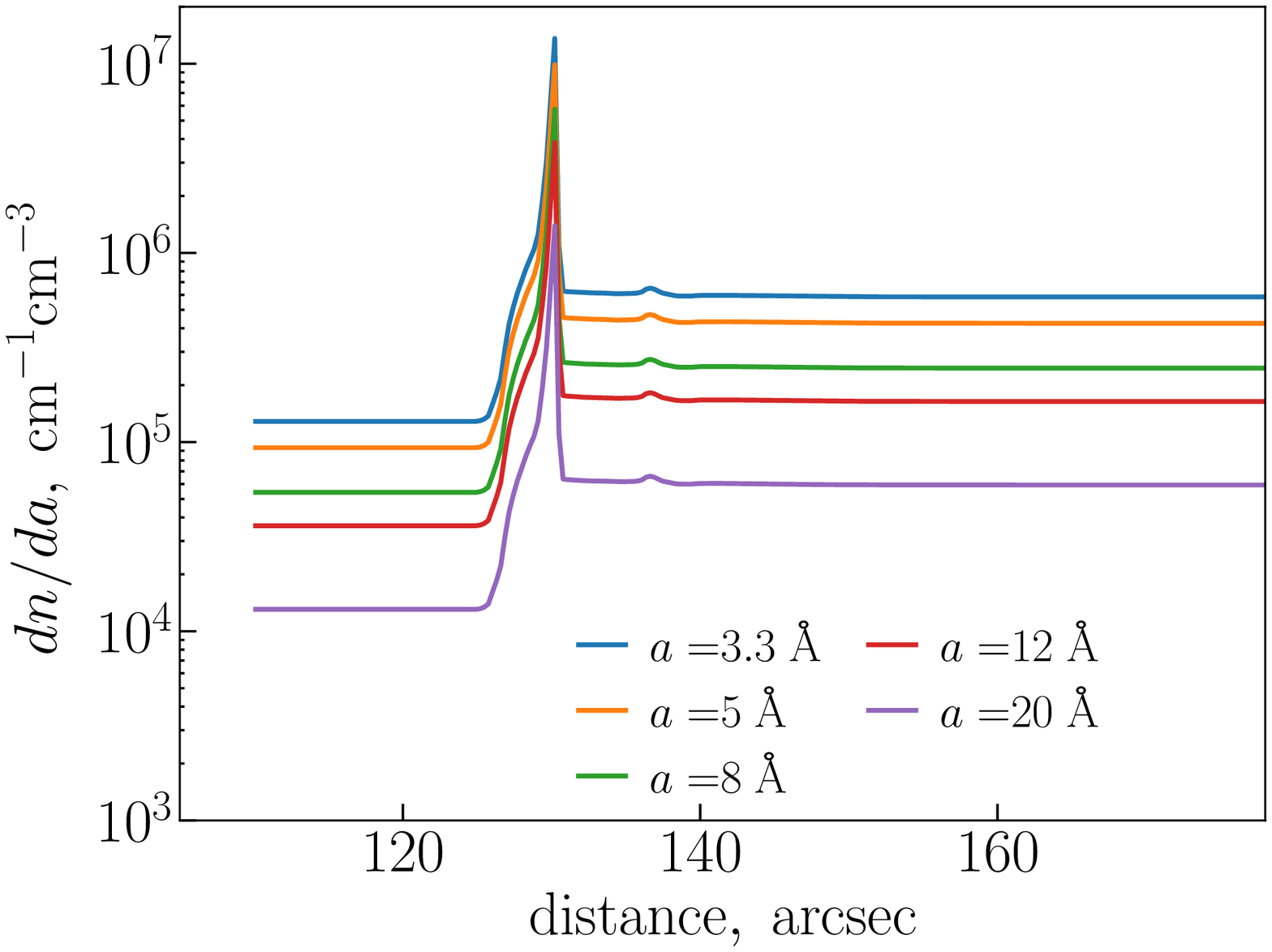}
\caption{Upper panels: physical parameters along the Orion Bar according to the {\tt MARION} model. Bottom left panel: parameters $\chi\sqrt{T_{\rm gas}}/n_{\rm e}$ and  $\chi/n_{\rm H}$ along the Orion Bar. Bottom right panel: values of the initial size distribution for some specific sizes along the Orion Bar.}
\label{fig: par}
\end{figure*}

That model is not intended to be an evolutionary model of the Orion Bar, but rather reproduces a `moving stationary' situation for the ionization and dissociation fronts advancing into molecular material. Parameter distributions shown in Fig.~\ref{fig: par} correspond to some evolutionary time of the model, when the overall structure of the studied region has already been established. This structure is kept fixed during the dust evolution computation. Using the model, we match distances expressed in values of extinction $A_{\rm V}$ and spatial distances in arcsec. We model the PAH evolution in the PDR at distances from 110 to 180~arcsec from $\theta^1$~Ori~C, where the IR features change sharply. We use gas temperature $T_{\rm gas}$, atomic hydrogen number density $n_{\rm H}$, number density of molecular hydrogen $n_{\rm H_2}$, $n(\rm H) = n_{\rm H} + n_{\rm H_2}$, radiation field intensity $\chi$, number density of electrons $n_{\rm e}$ and number density of ionized and atomic carbon ($n_{{\rm C}^{+}}$ and $n_{\rm C}$) from that model. 

This model region can be divided into three parts with different physical conditions: the \ion{H}{II} region, the PDR and the molecular cloud (MC). We add photons with energy exceeding 13.6~eV, and with photon flux corresponding to an O6-type star, in order to calculate the evolution of PAHs in the \ion{H}{II} region as \citet{kirsanova19} excluded these photons. In Fig.~\ref{fig: par}, we also show ratio $\chi\sqrt{T_{\rm gas}}/n_{\rm e}$, characterising the charge balance of PAHs and $\chi/n_{\rm H}$, which indicates the efficiency of losing or adding of H atoms \citep{tielens05}. 

The dust size distribution ($dn/da$) of \citet{kirsanova19} extends up to several microns. As we are interested only in PAH-size grains in the present study, we model the evolution of particles with sizes between 3.3~\AA{} and 25~\AA{} dividing this range into $N_{\rm m}=10$ bins. The initial values of the size distributions for some specific sizes are shown in Fig.~\ref{fig: par}~(bottom right panel). The range of hydrogenation levels is divided into $N_{\rm mH}=5$ bins, spanning a range from 0 to $2.5\cdot X_{\rm H}^{0}$ in $X_{\rm H}$. We follow the evolution of each bin, taking into account possible bin-to-bin exchanges. Finally, as mentioned above, we consider the evolution of PAHs at different ionization states. We solve the system of equations \eqref{main_eq} for PAHs with charge numbers from $-2$ to 8 ($N_Z=11$), as these are the most populated charge states of the PAHs in the conditions of the Orion Bar. 

The {\tt DustEm} tool \citep{dustem} is used to compute synthetic spectra in the mid-IR range. We consider $N_{\rm mH}\times N_Z=55$ PAH types, each type having its own size distribution obtained with the model. In addition, we include large graphite and silicate grains. We do not consider their evolution, but add them to the calculations of the synthetic spectra using the size distributions from \citetalias{DL07}.

The absorption cross-sections ($C_{\rm abs}$) of all dust types are adopted from \citetalias{DL07}, with modifications concerning the bands at 3.3, 3.4, and 11.2~$\mu$m for PAHs in the super-hydrogenated and dehydrogenated states. We describe these modifications in Appendices~\ref{sect: cabs} and \ref{sect: cabs2}. The NASA Ames PAH database is used as a source of parameters of IR bands. The total synthetic spectrum at any spatial point is obtained by summing the spectra of all dust components, i.e.
\begin{equation}
\begin{gathered}
     I_{\rm total}(\nu) = \sum\limits_{i=1}^{N_{\rm mH}\times N_{Z}+2}f_{i}(Z)\int\frac{dn_i}{da}da \times \\ 
     \times \int C_{\rm abs}(i, a,Z,\nu)B_{\nu}(T)\frac{dP}{dT}(i,a)dT,
     \end{gathered}
\end{equation}
where $B_{\nu}(T)$ is the Planck function, and $dP/dT$ is the temperature distribution for each grain at a certain computational cell.

We applied the same procedure as in Sects.~\ref{Spitzer spectroscopic data} and \ref{UKIRT spectroscopic data} to obtain the band intensities at 3.3, 3.4 and 11.2~$\mu$m. For the bands at 3.6, 6.6, and 7.7~$\mu$m, we use the formulae presented in the IRAC and FORCAST descriptions \citep{hora08, herter13}, which can be expressed as follows
\begin{equation}
    I_{\rm band} = \frac{\int R_{\rm band}(\nu)I_{\rm total}(\nu)d\nu}{\int(\nu/\nu_{\rm band})^{-1}R_{\rm band}d\nu},
\end{equation}
where $R_{\rm band}$ is the system response function related with the response function $S_{\rm band}(\nu)$ as $\propto(1/h\nu)S_{\rm band}(\nu)$. The response functions were obtained from the IRAC instrumentation handbook\footnote{\url{https://irsa.ipac.caltech.edu/data/SPITZER/docs/irac/}} and the FORCAST instrument description\footnote{\url{https://www.sofia.usra.edu/science/instruments/forcast}}. Note that we simulate the band intensities, while we measure flux densities using the observational data. As these two values differ by a constant factor, we assume that their ratios can be compared.

There is a faint background emission near 3~$\mu$m at the ISO and UKIRT spectra. Since we measure the intensity at 3.6~$\mu$m from photometric image, this background emission remains in our measurements, and we cannot estimate it accurately. We consider that this background is related with the radiation field (e.g. via light scattering) and varies with its intensity. Therefore we roughly approximate this background in the following way: we calculate the background flux density from the ISO spectrum and the fraction of the background in the 3.6~$\mu$m-band flux density. Then we add this additional fraction to the simulated 3.6~$\mu$m-band intensities, scaling it by the ratio between $\chi$ at a current distance and $\chi$ at the ISO slit location. The intensity of the 3.6~$\mu$m-band with the added background is designated as $I_{3.6}^{\rm bkg}$.

\section{Results of the model}\label{sec:modelresults}

In this section, we present results of our calculations and compare them with the observational data. We show how the PAH distributions over sizes and ionization/hydrogenation states vary, depending on a distance from an ionizing source, and also present the dependence of band intensity ratios on the distance. We present results for representative locations within the \ion{H}{II} region, the PDR (two points designated by numbers 1 and 2), DF, and MC. The distances from the ionizing source to these locations are 110, 120, 129, 130, and 160~arcsec, respectively (see crosses in Fig.~\ref{fig: par}, upper left).

\subsection{PAH size distribution}

\begin{figure*}
\includegraphics[width=0.49\textwidth]{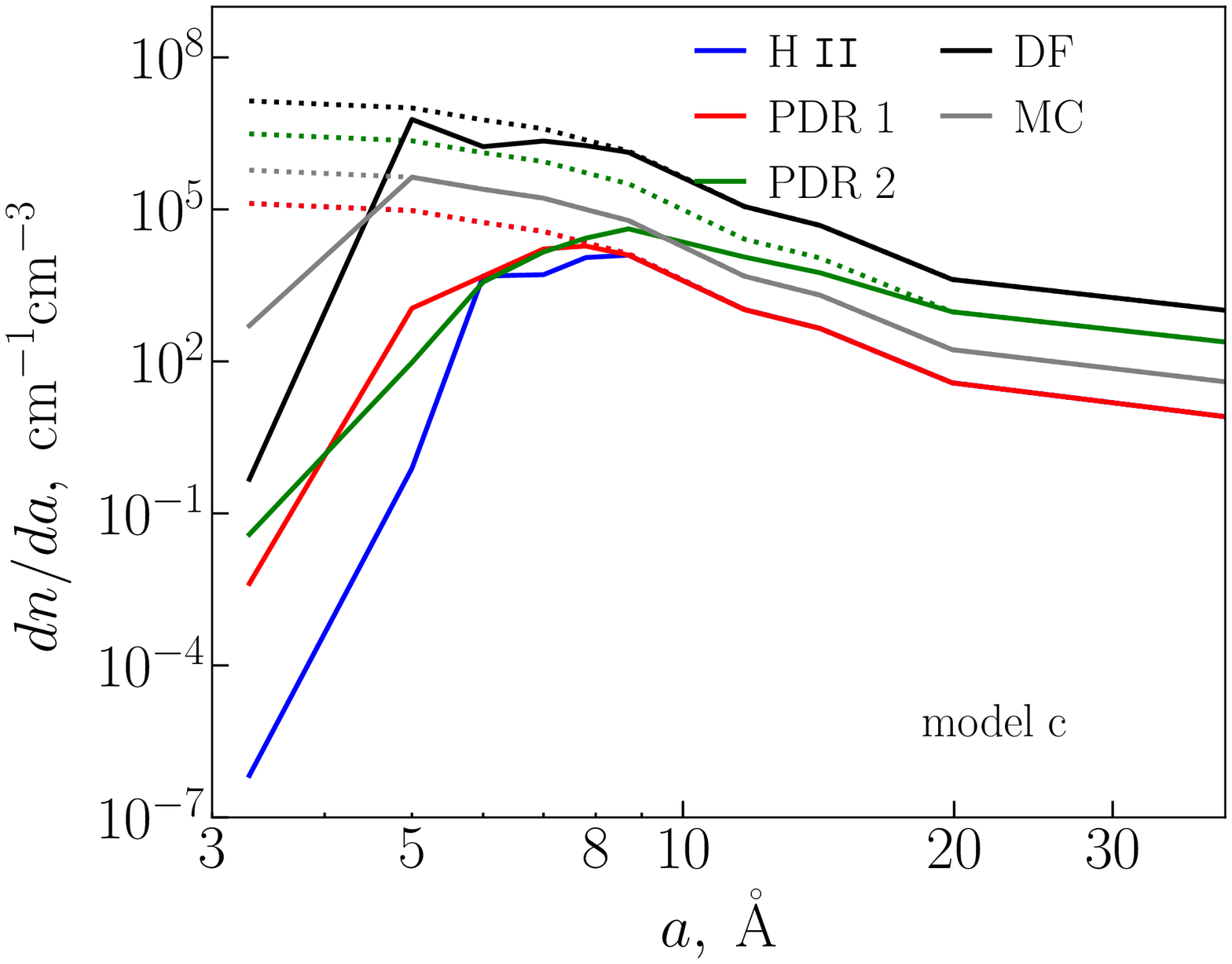}
\includegraphics[width=0.49\textwidth]{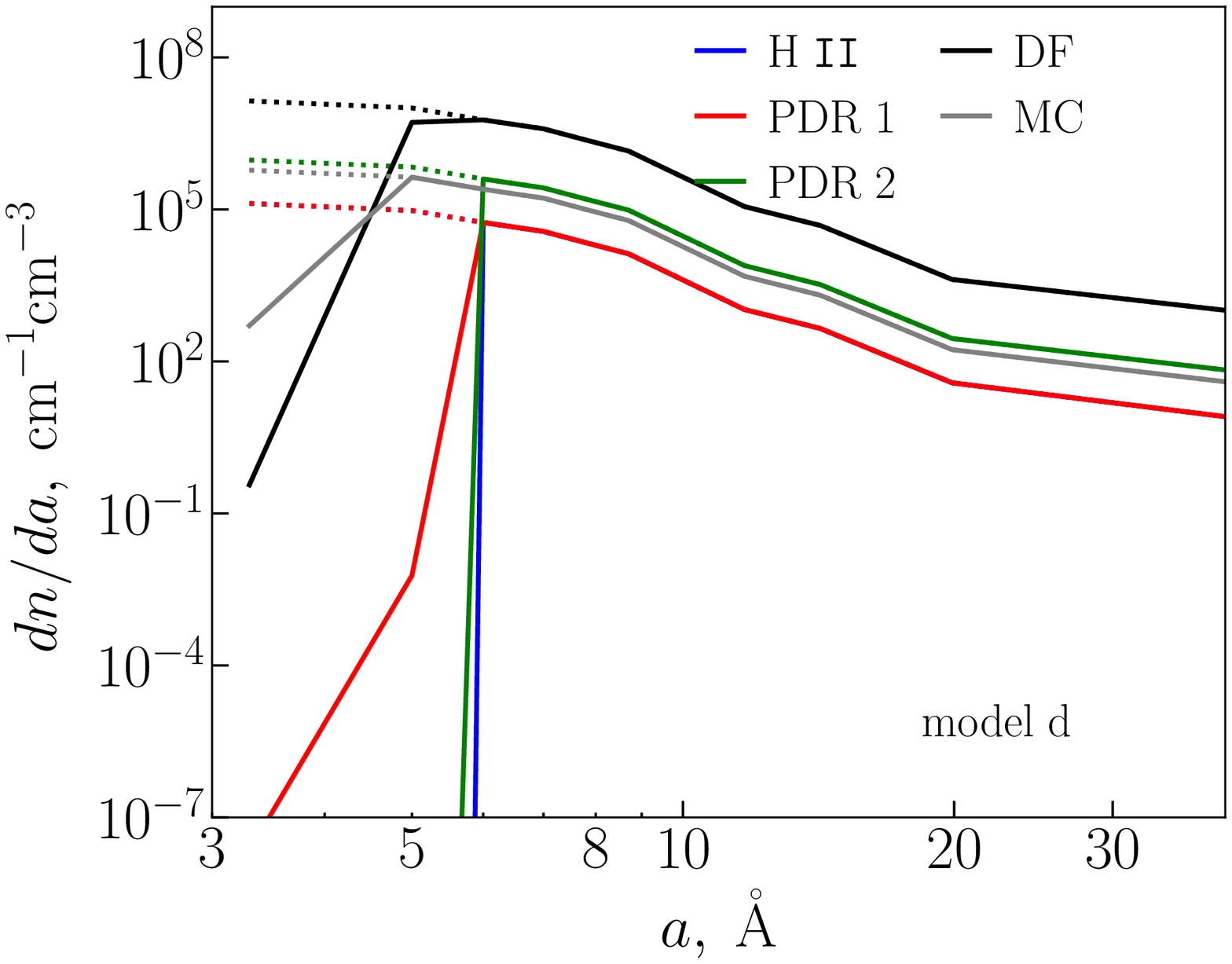}
\caption{PAH size distribution in the positions corresponding to \ion{H}{II}, PDR~1, PDR~2, DF, and MC locations. Dotted lines of the corresponding colour show the initial distribution.}
\label{fig: size_distr}
\end{figure*}

The initial dust size distributions and results of our calculations at each location are shown in Fig.~\ref{fig: size_distr}. The size distribution does not change when no loss or only H loss may occur (models `a' and `b'), therefore, we only present results for models `c' (Fig.~\ref{fig: size_distr}, left) and `d' (Fig.~\ref{fig: size_distr}, right) in this subsection. We demonstrate the size distributions of PAHs in \ion{H}{II}, PDR~1, PDR~2, DF, and MC locations of the model region after $10^5$~yr, which is assumed to be approximate age of the Orion Bar PDR \citep{salgado16, kirsanova19}. 

The critical size of PAHs in model `c', below which they are destroyed, changes from one position to another. This size is  8~\AA{} at the \ion{H}{II}, PDR~1, and DF points, $\approx$~15~\AA{} at PDR~2 and $\approx5$~\AA{} at MC. From the size distributions, we note that the fraction of the destroyed PAHs is generally larger for smaller sizes, and that the most efficient destruction of PAHs occurs in PDR~2.

In model `d', where the C loss of large PAHs ($N_{\rm C}>60$) is restricted, the abundances of PAHs with $N_{\rm C}<60$ goes to zero at the \ion{H}{II}, PDR~1 and PDR~2 locations. All distributions are similar to the ones obtained in model `c', except that the abundances of the smallest grains are lower in PDR~1 and PDR~2 in case `d', as large PAHs are not destroyed, and their fragments do not pass into the bins of smaller sizes. 

We note that regardless of which model is used, PAHs with radii smaller than 5~\AA{} are destroyed throughout the model region, except for the deep interior of the MC (i.e., farther than 160~arcsec from the star). Thus such small PAHs may only survive in the shielded regions of the MC.

\begin{figure*}
\includegraphics[width=0.49\textwidth]{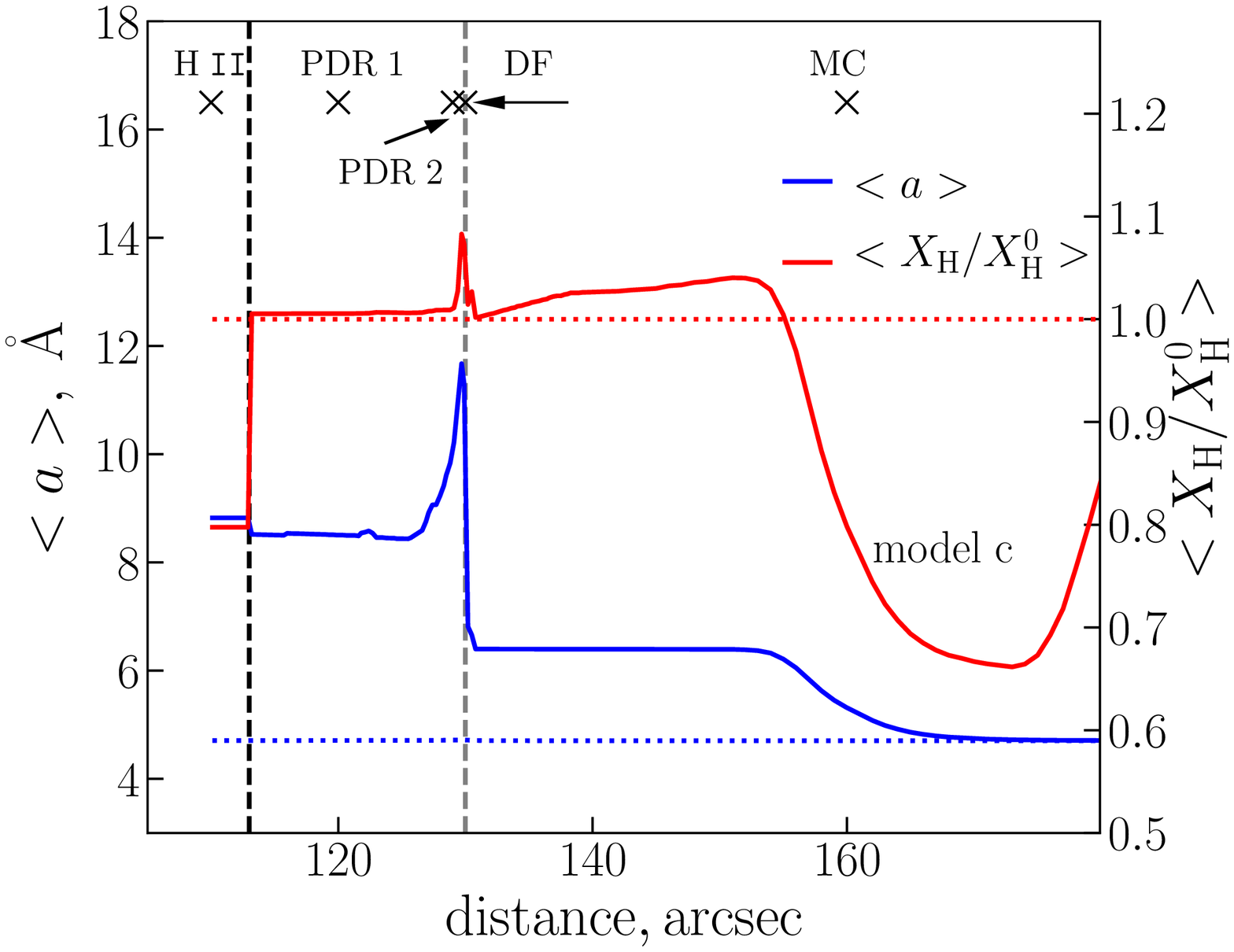}
\includegraphics[width=0.49\textwidth]{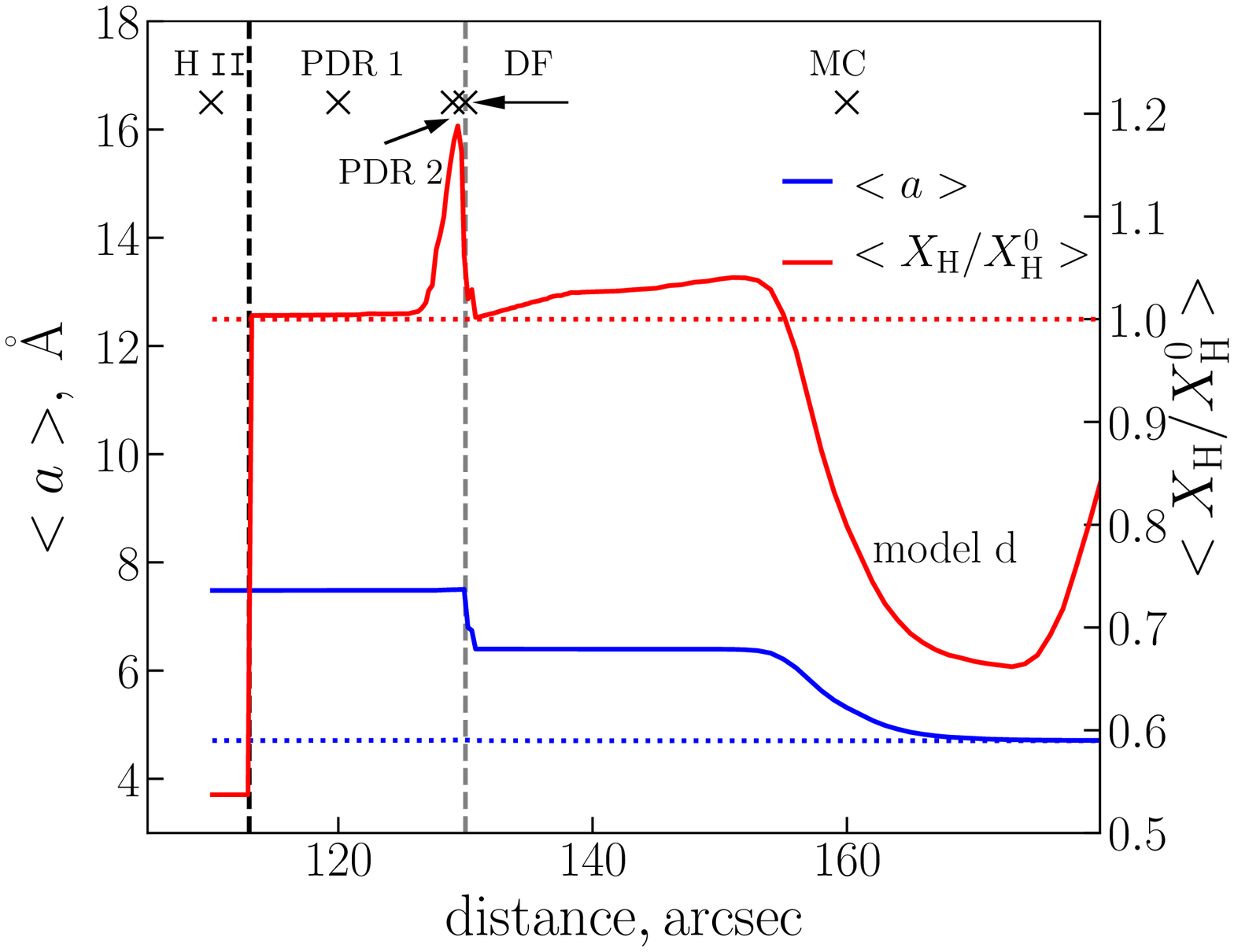}
\caption{Weighted average PAH size ($\langle a\rangle$, blue line) and hydrogenation level relative to the initial one ($\langle X_{\rm H}/X_{\rm H}^{0}\rangle$, red line) along the Orion Bar. The initial values are shown by dotted lines of the corresponding colour.}
\label{fig: size_av}
\end{figure*}

Another way to trace the destruction of PAHs at various locations is to estimate the average size of the PAHs ($\langle a\rangle$) across the object. We calculate it as a weighted average value over all size bins. A weight of a size bin is taken to be a sum of number densities of all hydrogenation and ionization states within the bin. We present the radial profiles of the PAH average radius in Fig.~\ref{fig: size_av} (blue line). In model `c', the average size of PAHs at the IF is approximately 8~\AA{}. At $\sim125$~arcsec it starts to increase, reaching 12~\AA{} at $\approx129$~arcsec. Then $\langle a\rangle$ rapidly drops to $\sim6$~\AA{} at $\sim130$~arcsec (location of the DF) and smoothly decreases down to the minimum size in the MC. This behaviour of the average PAH size demonstrates that within the framework of model `c' the most adverse environment for PAHs in the model region is PDR~2, where the hydrogen number density has its maximum. The high number density of atomic hydrogen in PDR~2 leads to efficient hydrogenation of PAHs. This is especially true for neutral PAHs, which are highly reactive with hydrogen at high temperature \citep{rauls08}, although ionized PAHs also efficiently interact with hydrogen atoms, if $n(\rm H)$ is high. As PAHs in the super-hydrogenated state are unstable and the radiation field intensity is still quite high in this area ($\chi>10^2$), these PAHs are destroyed quickly. The $\langle a\rangle$ value does not rise at the PDR~2 location in model `d' due to the selected size limit against the destruction by UV~radiation.

\subsection{Ionization and hydrogenation state of PAHs}

A red line in Fig.~\ref{fig: size_av} shows the dependence of the weighted average hydrogenation state ($X_{\rm H}/X_{\rm H}^{0}$, see Sec.~\ref{sect: model}) on a distance for the models `c' and `d'. The hydrogenation level slightly exceeds 1 at the PDR region up to $\sim125$\arcsec, but farther out the fraction of H atoms in PAHs grows. The highest hydrogenation level (1.1 and 1.2 for models `c' and `d', respectively) appears in the PDR~2 location and coincides with the highest $n_{\rm H}$ value. In the MC, it drops to 1--1.05 and then decreases to 0.6 behind $\approx 150$\arcsec. The smallest PAHs start to survive deeply in MC (no C loss), but they are still dehydrogenated at the same time (see details below). These PAHs are getting stable even to H loss, and the hydrogenation level increases again behind 170~arcsec . 

\begin{figure*}
	\includegraphics[width=0.95\textwidth]{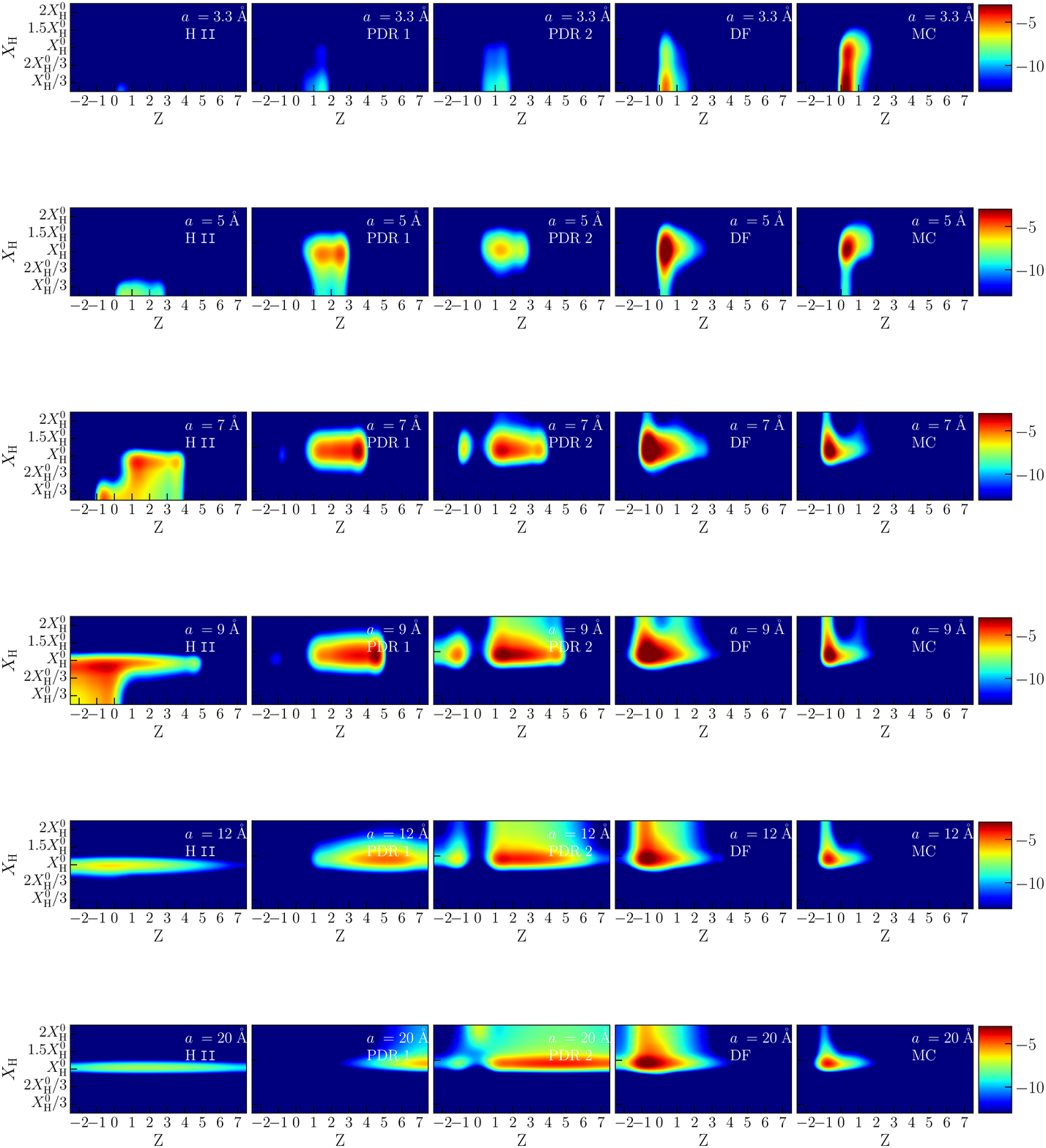}
	\caption{Maps of PAH hydrogenation and charge states (model `c'). Columns correspond to the positions of \ion{H}{II}, PDR~1, PDR~2, DF, and MC. Rows correspond to different sizes (3.3, 5, 7, 9, 12, 20~\AA{}).}
	\label{fig: ion_hydr_map}
\end{figure*}

We show colour maps illustrating the distribution of states for model `c' in Fig.~\ref{fig: ion_hydr_map}, where colour indicates the number density of PAHs in a specific state. Results of models `c' and `d' are qualitatively similar. These maps allow assessing the states of specific PAHs (six size bins from 3.3 to 20~\AA{}) under certain conditions and number densities of PAHs at the selected locations. We note again that all dust particles are mostly abundant initially at the DF, according to the {\tt MARION} model, but their abundances change due to evolution under the UV~radiation.

The smallest PAHs (3.3~~\AA{}, top row in Fig.~\ref{fig: ion_hydr_map}) are nearly absent in \ion{H}{II}. Their number density grows from PDR~1 and reaches a maximum in MC. These PAHs survive only in the MC. The charge number of these PAHs is mostly 1 in PDR and 0 in DF and MC. They are mostly dehydrogenated nearly everywhere in the Orion Bar, except for the deep interior of the MC.

The number density of 5~\AA{} PAHs (second row in Fig.~\ref{fig: ion_hydr_map}) is smaller by several orders of magnitude in the \ion{H}{II}, PDR~1 and PDR~2 relative to the DF and MC locations. Their charge ranges from 0 to 3 in the PDR locations and at the DF, and from 0 to 2 in the MC. In PDR~1 and PDR~2, the number density peaks in ionized states, while neutral PAHs are most abundant at the DF and in the MC. Unlike 3~\AA{} PAHs, these particles are mostly in the normal hydrogenation state, though a small fraction of dehydrogenated PAHs also exist. 
 
The difference in number densities of the 7~\AA{} PAHs at the selected locations is not as prominent as for the smaller sizes (third row in Fig.~\ref{fig: ion_hydr_map}). These PAHs may lose their carbon atoms only in the PDR and at the DF, but less efficiently than smaller particles. The range of charges is wider: in \ion{H}{II} and PDR the charge varies from --1 to 4, at the DF and in the MC it varies from --1 to 3. The dominant charge state shifts from a positive charge in PDR~1 and PDR~2 to a negative charge at DF and in MC. These PAHs are mostly in the normal hydrogenated state, with the insignificant fraction being super-hydrogenated. We emphasise a gap indicating a deficit of neutral PAHs of this size in the PDR locations, though both anions and cations exist there. The abundance of neutral PAHs decreases due to the rapid hydrogenation and subsequent destruction. The same feature appears also for larger sizes. 

The PAHs with radius of 9~\AA{} (fourth row in Fig.~\ref{fig: ion_hydr_map}) can be destroyed only in PDR~2, but not substantially. The range of PAH charges is from --2 to 5 in PDR, from --2 to 4 at DF, and from --1 to 2 in MC. In PDR~1 and PDR~2 the PAHs are positively charged, and they become anions in DF and in MC. There are no neutral PAHs in PDR~1 and PDR~2. These PAHs stay in the normal hydrogenation state mostly, but some PAHs are super-hydrogenated at all locations. The fraction of H-PAHs becomes especially large in PDR~2 and DF.

The maps for the PAHs of radius 12~\AA{} (fifth row in Fig.~\ref{fig: ion_hydr_map}) look similar to the maps of the previous size, but they are even more hydrogenated in PDR and DF even though they are mainly in the normal hydrogenation state. There are no neutral PAHs in PDR~1 and PDR~2 due to photo-destruction as well as for two previous sizes. The fraction of neutral PAHs is insignificant in PDR~1, therefore their destruction does not contribute to the PAH size distribution in Fig.~\ref{fig: size_distr}, while their fraction in PDR~2 is more considerable and their destruction is reflected on the size distribution. The range of charges is wide in PDR~1 and 2 (from --2 to 10), but becomes narrower in DF (from --2 to 4) and MC (from --2 to 3). There is a shift from positively charged to negatively charged particles after transition from PDR~1 and PDR~2 to DF and MC.

Finally, we present the maps for 20~\AA{} PAHs (bottom row in Fig.~\ref{fig: ion_hydr_map}) though these large grains are practically insensitive to the UV~radiation. They are mostly positively charged in PDR~1 and PDR~2 and negatively charged in the DF and MC locations. As for the hydrogenation state, their behaviour is similar to the behaviour of smaller PAHs. Some of them are super-hydrogenated, but they are mostly in the normal hydrogenated state. Also, even neutral PAHs of this size, which can be quickly hydrogenated, survive because these PAHs are stable against photo-destruction even in the super-hydrogenated state.

\subsection{Comparison between the observed and simulated PAH emission}\label{sect: comparison}

\begin{figure*}
	\includegraphics[width=1.0\textwidth]{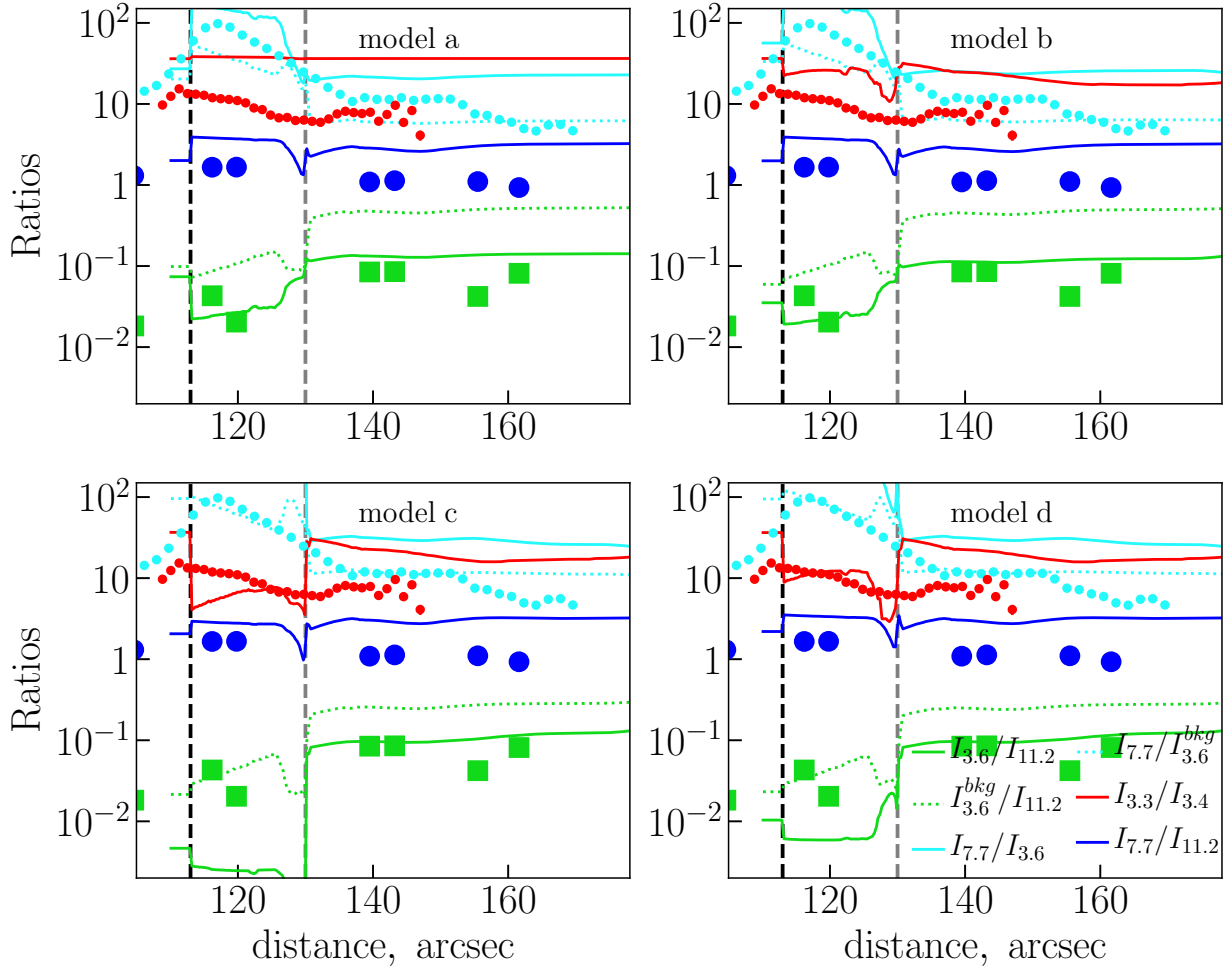}
	\caption{Ratios of the band intensities obtained from the synthetic spectra. The model cases (`a', `b', `c', `d') are marked in corresponding plots. Observational data are also shown, as in Fig.~\ref{fig: sum_obs}.}
	\label{fig: ratios}
\end{figure*}

In Fig.~\ref{fig: ratios} we show the observed and simulated ratios $I_{3.6}/I_{11.2}$ and $I^{\rm bkg}_{3.6}/I_{11.2}$ (PAH size), $I_{3.3}/I_{3.4}$ (ratio of aliphatic to aromatic groups), $I_{7.7}/I_{3.6}$ (PAH size and charge) and $I_{7.7}/I_{11.2}$ (PAH charge). The colours in the figure repeat the colours used for the observational the data in Fig.~\ref{fig: sum_obs}.

\subsubsection{$F_{3.6}/F_{11.2}$ ratio}

The observed ratio $F_{3.6}/F_{11.2}$ (green squares) has a minimum at the IF and then stays nearly constant $\approx 0.1$ in the MC. The ratio anti-correlates with parameters $\chi \sqrt{T_{\rm gas}}/n_{\rm e}$ and  $\chi/n_{\rm H}$ (compare with Fig.~\ref{fig: par}). We find a trough in the radial profile of the simulated $F_{3.6}/F_{11.2}$ ratio (green lines) between the IF and DF in all models. If we `switch on' carbon skeleton destruction and consider models `c' and `d', the trough is deeper, and its depth gets larger if destruction of larger grains is allowed (model `c'). A quick look at observed and computed $F_{3.6}/F_{11.2}$ ($I_{3.6}/I_{11.2}$) ratios leads to conclusion that only models without PAH destruction (`a' and `b') are consistent with the observational data. But if we consider simulated $I_{3.6}^{\rm bkg}$ instead of $I_{3.6}$, then the ratio becomes higher than the observational one in the models without C loss up to half order of magnitude, but the ratio is consistent with the observations in the models `c' and `d', where PAH carbon skeleton can be destroyed. We know that the background emission exists, therefore we choose the $I_{3.6}^{\rm bkg}$  values and prefer the models with C loss to fit the $F_{3.6}/F_{11.2}$ ratio.

In our calculations without restrictions on the PAH size, the destruction occurs quickly in PDR~1 and PDR~2 due to the harsh radiation field, particularly the destruction is efficient in the PDR~2 due to the rapid hydrogenation of PAHs at this location and the low stability of H-PAHs. Such PAHs disappear at these locations within the time interval shorter than the model age of the Orion Bar PDR. The PAHs up to $a\approx15$~\AA{} are destroyed (see Fig.~\ref{fig: size_av}), and their fragments can partially refill the bins of smaller sizes. The average size has a maximum in the PDR~2. The destruction leads to the significant weakening of the emission at 3.6~$\mu$m, and the ratio $I_{3.6}/I_{11.2}$ becomes too low. That is why we also checked the cases with the restriction of the destroying PAHs by $N_{\rm C}=60$ (`d') and restriction for destruction of carbon skeleton at all (`b'). In the case `d', the average size does not drastically change, and there is no size growth as we move from the IF to the PDR locations. Thus, we conclude that model `d' fit the observational $I_{3.6}/I_{11.2}$  ratio better than other models. We also mention that if the C loss is allowed then PAHs of small sizes ($a<5$~\AA{}) may exist in the Orion Bar only deep in MC. 

\subsubsection{$F_{7.7}/F_{3.6}$ ratio}

The observed $F_{7.7}^{\rm conv}/F_{3.6}$ ratio (cyan dashed lines) is higher than the simulated one in PDR for all considered models by 0.5--1 order of magnitude or ever more. The simulated ratio becomes closer to the observed one (within 0.5 order of magnitude) in MC region of models `a' and `b'. However, if we again consider the $I_{7.7}/I_{3.6}^{\rm bkg}$ ratio, models `b' and `d' satisfactorily describe the PDR region and the MC zone.

\subsubsection{$F_{7.7}/F_{11.2}$ ratio}

The simulated $I_{7.7}/I_{11.2}$ ratio (blue lines) looks similar in all models (blue circles). The ratio decreases from the IF to the MC both in observations and in the model, but the simulated values are 2--3 times (sometimes up to $\approx$~order of magnitude) higher than the observational points. There might be two reasons for that: 1) the deficit of neutral PAHs both in the PDR and the MC in our calculations; 2) the necessity to use optical properties determined specifically for negatively charged PAHs. 

The first reason is related to the adopted ionization potential, which is taken from \cite{wd01_charge}. It is a generalised potential that suits some `average' PAHs. However, certain deviations are possible for individual PAHs, so that PAHs with other potentials or even neutral PAHs can exist in PDR and MC zones. In the latter case the 11.2~$\mu$m-band will be enhanced, and therefore the ratio $I_{7.7}/I_{11.2}$ will be smaller. According to our charge calculations, some neutral PAHs do exist near the DF, and at this point the ratio $I_{7.7}/I_{11.2}$ is around observational points (models `b', `c', `d'). This result correlates with the results of \cite{compiegne07}, who showed that neutral PAHs are needed to explain the intensity of the 11.3~$\mu$m-band in the Horsehead nebula PDR.

The second reason is our assumption on the identical optical properties for negatively and positively charged PAHs, based on recent investigations of PAH spectra \citep{gao14, burago18}. Thus, while ionized PAHs in the MC are dominated by negatively charged PAHs, we treat them as positively charged PAHs instead, computing the synthetic spectra. Currently, no optical properties are available that would be computed or measured specifically for anions, but if in fact optical properties of positively and negatively charged PAHs differ significantly, the synthetic ratio $I_{7.7}/I_{11.2}$ computed with more appropriate optical properties, may be closer to the observed values in the MC.

\subsubsection{$F_{3.3}/F_{3.4}$ ratio}

The observed $F_{3.3}/F_{3.4}$ ratio (red circles) drops by at least a factor of 2 from the IF to the DF and stops decreasing after the DF. In the MC the ratio fluctuates, staying roughly at the same level. None of our models reproduce this behaviour completely, but in model `d' the simulated $I_{3.3}/I_{3.4}$ ratio (red lines) follows the observed radial profile closely in the IF and PDR~1 locations (within several percent). If PAHs cannot be destroyed (model `b'), then the 3.3~$\mu$m band flux would be large compared to the 3.4~$\mu$m band flux. This is because as the small PAHs ($a\leq5$~\AA{}) are dehydrogenated in the PDR, but their number density is the largest of all the PAHs. When small PAHs are destroyed (model `c'), the 3.4~$\mu$m band flux increases relative to the 3.3~$\mu$m band flux, as the large PAHs can be super-hydrogenated. Therefore, model `d' is an intermediate case when the relation between the abundances of dehydrogenated small PAHs and large H-PAHs is consistent with the observations.

However, none of our proposed models are able to adequately describe the observations in the MC. A direct interpretation of the observed ratio suggests that PAHs may be super-hydrogenated in the MC. This could be the case, if the PAHs are able to acquire hydrogen atoms from H$_2$ molecules. In our model, PAHs do not interact with molecular hydrogen, and the density of atomic hydrogen is too low to affect the abundance of H-PAHs. This behaviour could also be related to the non-zero fraction of atomic hydrogen in the molecular gas of the Orion Bar PDR. Moreover, the smallest PAHs, which are the most abundant in the MC, are still dehydrogenated. Consequently, even model `d' overestimates the $F_{3.3}/F_{3.4}$ ratio in the MC. The depth of the trough near PDR~2 varies between models, with models `c' and `d' having the maximum depth.

\section{Discussion}\label{sect: discus}

\subsection{Comparison with other studies of the Orion Bar}

Many works have been devoted to PAH studies in the Orion Bar. Firstly the unidentified infrared emission features at 8--13~$\mu$m in this object were discovered by \cite{aitken79}. The emission was found to peak in the neutral region of the PDR, behind the IF, where radiation gets weaker. Since \cite{leger84} suggested PAHs to be responsible for unidentified infrared emission features, this hypothesis was almost unanimously adopted for the mid-IR emission in this object as well. \citet{bregman89, roche89, giard94, geballe89, allain96_1} proposed that PAHs can hardly survive in the \ion{H}{II} region, and this is why the peak of their emission is shifted to the neutral region. However \cite{kassis06} adapted a simple attenuation model to the region and showed that the PAH emission reflects the geometry of this object. Namely, the sharp increasing of the PAH column density behind the IF leads to increasing of mid-IR emission.

Based on the PAH fluorescent model \cite{schutte93, bregman94} summarised that PAHs have $N_{\rm C}$ from less than 80 to 10$^2-10^5$, and PAHs are rather not dehydrogenated except the smallest ones with $N_{\rm C}<24$. \cite{allamandola99, bakes01a, bakes01b} modelled synthesised spectra of mixtures of particular PAHs 
and found that large PAH cations ($N_{\rm C}\approx50-100$, singly and multiply charged) dominate at the surface of the PDR at $A_{\rm V}\approx0.1$, and the structure of the PAHs is rather symmetrical and condensed. These PAHs are thermodynamically stable and likely survive under the harsh conditions. Deeper to the more attenuated region the PAH mixture is dominated by neutral and negatively charged PAHs while their size distribution is shifted to small PAHs. On the other hand, \cite{kassis06} concluded that there is no clear transition from positively charged to neutral and negatively charged PAHs from bright to faint region in this object. Instead of that PAHs rather change their other characteristics like molecular structure, size or chemical composition.

Recently \cite{knight21} demonstrated observational profiles of ratios $F_{11.2}/F_{3.3}$ and $F_{8.0}/F_{11.2}$. The peculiar profile of the ratio $F_{8.0}/F_{11.2}$ was presented where the ratio has a minimum closer to the ionizing source. This behaviour is unexpected as the ratio should be higher where the PAHs are ionized and lower under the mild conditions of molecular cloud. They conclude that the ratio $F_{8.0}/F_{11.2}$ does not strictly reflect the ionization of PAHs and is smudged by their dehydrogenation. The ratio $F_{11.2}/F_{3.3}$ has a maximum at the bright region of the PDR decreasing to the MC which is explained by the photo-destruction of small PAHs in the former region. However, they emphasise that there is not significant variation of the ratio in the MC where the average PAH size is estimated around 70--85 carbon atoms.

The results of the works above prove that evolutionary processes of PAHs should be considered for detailed modelling of mid-IR emission in this object. However, there was no detailed model that could describe charge, size and the hydrogenation state of PAHs simultaneously. Our results based on such a model and observations generally confirm the conclusions of the previous works, and also allow us to reconstruct the PAH parameters in the Orion~Bar PDR in detail. We show that large PAHs dominate over neutral and negatively charged PAHs in the PDR while the opposite situation is in the MC. However, our model overestimates the ratio $F_{7.7}/F_{11.2}$, which hints at additional factors, beside of charge, that influence the ratio. \cite{knight21} suggested dehydrogenation as one of these factors, but we do not see prevailing of dehydrogenation in the PDR. Unlike \cite{knight21}, we do not see a peculiar behaviour of $F_{7.7}/F_{11.2}$. The possible reason is that we use different bands and instruments, though the ratio is expected to be similar. However, we note the profiles of our $F_{3.6}/F_{11.2}$ and their $F_{3.3}/F_{11.2}$ are consistent with each other.

Our synthetic band intensity ratios are reasonably close to the observations (up to a factor of 3). Our model (in particular, case `d'), ISO spectrum fitting and size estimations with using the diagrams from \cite{marag20} indicate that the PAHs should include around 100 carbon atoms or even more in the PDR. Thus, our results from different methods are consistent with each other. Also, they are generally consistent with estimates from previous  works~\citep{bakes01a, kassis06, knight21}. Unlike the previous works, our modelling naturally shows the processes that lead to the observed distribution of PAH sizes, charges and hydrogenation states in the Orion Bar PDR.

\subsection{Origin of the 3.4~$\mu$m band}

It is worth discussing separately the origin of the 3.4~$\mu$m band that has been intensively discussed in the literature concerning different objects, including the Orion Bar. Besides the H-PAHs considered in the present work, a number of other carriers has been suggested: evaporating PAH clusters,  methylated PAHs (Me-PAHs), HACs, mixed aromatic-aliphatic organic nanoparticles (MAONs), and the anharmonicity effect~\citep{bakes01b, kwok11, ld12, yang13, jones13, sadjadi15, molpeceres17, chen18a}. However, precise calculations of the evolution and IR emission of these particles under the conditions of the Orion Bar have not yet been performed. The H-PAHs, Me-PAHs, HACs, MAONs seem to be unstable under UV radiation, as are H-PAHs \citep{mennella01, jones13, rapacioli18}. They may be destroyed quickly, or at least dehydrogenated, in the harsh radiation field of the Orion Bar, losing the ability to emit at 3.4~$\mu$m. In this work, we quantitatively analysed whether H-PAHs can exist and emit sufficiently in the Orion Bar or not. In our calculations, the hydrogenation process of small PAHs is compensated by the H loss process due to photodestruction, but PAHs larger than 5--7~\AA{} can be super-hydrogenated, and thus contribute to the 3.4~$\mu$m band enough to explain its flux relative to the 3.3~$\mu$m band. Therefore, the 3.4~$\mu$m band can be related to aliphatic bonds in H-PAHs. The hypothesis was discussed earlier in \cite{schutte93, sloan97}, and here we approve the plausibility of the suggestion, although other carriers should not be excluded. 

\subsection{Problems with the ISO spectrum fitting}

One problem with the ISO spectrum fitting is the shift of the observed 6.2~$\mu$m band and fitted 6.3~$\mu$m band. \citet{bauschlicher08} showed that the 6.2~$\mu$m feature could originate at 6.3~$\mu$m, but the inclusion of nitrogen atoms in the PAHs shifts the feature to 6.2~$\mu$m. The database we use for the fitting contains a small number of large ($N>80$) PAHs with nitrogen atoms, therefore the lack of data for nitrogenated PAHs may explain the shift.

Another problem appears at 12--14~$\mu$m, where our PAH mixture underestimates the mid-IR flux, and overestimates the flux around 6--7~$\mu$m. This problem may be related to the high intensity of the UV field in the Orion Bar PDR, so that PAHs might reach energies equivalent to $\sim 10^3$~K. These hot PAHs contribute to the spectrum region near 6--7~$\mu$m, rather than to the region at 12--14~$\mu$m, which may explain why the feature at 6.2~$\mu$m is enhanced, while the features at 12--14~$\mu$m are reduced. 

We find that the most abundant species in the PAH mixture in our fit do not coincide with those from either of the lists presented by \cite{andrews15}. Only two PAHs (C$_{98}$H$_{28}^{+}$, C$_{112}$H$_{26}$), which play a minor role, are included to the fitting mixtures of both our work and the work of \cite{andrews15}. Taking the PAHs from the lists of \cite{andrews15} and calculating spectra with them under the conditions of the Orion Bar, we are not able to reproduce the 6.2 and 12--14~$\mu$m features of the ISO spectrum. Only emission of large PAHs, like those presented in Table~\ref{tab: fit}, satisfies the observed ratios between the 6.2, 7.7 and 11.2~$\mu$m features. However, these PAHs do not have strong features at 12--14~$\mu$m, so this range cannot be well fitted simultaneously with the other bands. The consideration of accurate heat capacities and cooling functions of individual PAHs might improve this fitting, while our approach, which uses the average PAH properties, makes the fit less reliable.

\cite{rosenberg14} demonstrated that a random PAH mixture can describe any astronomical mid-IR spectrum in the region of 5--15~$\mu$m. Our results are not consistent with their conclusions, since we show that only very large PAHs can reproduce the spectrum of the Orion Bar. Even then, the synthetic spectrum is still not satisfactorily close to the observed spectrum. In addition, some bands appear in the synthetic spectrum at wavelengths where no bands are observed (6.5--7, 9--10~$\mu$m). 

Based on the results of our fit to the ISO spectrum, the PAH mixture is mostly comprised of large PAHs (C$_{96}$H$_{26}$, C$_{98}$H$_{28}^{+}$, C$_{142}$H$_{30}$, etc.). However, there should also be a number of small dPAHs and normally hydrogenated PAHs (C$_{21}$, C$_{34}$H$_{20}^{+}$). The small PAHs affect the spectrum in some very specific features, and are necessary to fit the spectrum. Based on our models `c' and `d', the PAHs smaller than 6~\AA{} nearly disappear at the DF and in the PDR. In order to reproduce these small PAHs, there should be a way to produce small PAHs in these regions. 

The experiments of \citet{jochims94, zhen14, zhen15} showed that small PAHs may be destroyed efficiently under the UV radiation, while large PAHs can be ionized or lose hydrogen atoms. However, some experiments \citep[e.g.][]{ekern98, useli10, zhen15} also show that coronene (C$_{24}$H$_{12}$) has a very stable carbon skeleton and may lose only hydrogen atoms, and some other small PAHs (e.g. azulene) were unexpectedly discovered to be stable. \citet{ekern98} also showed that some PAHs, after losing just a few acetylene molecules, change their molecular structure and become photostable. In other words, the photodestruction of PAHs is complex, and some PAHs may be unexpected stable.

Another possibility is that the small PAHs appear in the PDR due to fragmentation of PAH clusters, or even HAC particles, however checking this hypothesis requires intensive calculations and experiments. Several efforts have been made \citep{rapacioli06, montillaud14, duley15}, but the question still requires additional investigation. Even if some small PAHs can be produced due to the fragmentation of larger complexes, these PAHs might be quickly destroyed, i.e. their production rate should be higher than their destruction rate.

\section{Summary and conclusions}
\label{sect: summary}

In this work, we aim to understand whether our PAH evolution model, which predicts the characteristics of PAHs under certain conditions, can explain the mid-IR observations of the Orion Bar PDR. We collected archival observational data and analysed them. We modelled the PAH photo-destruction under a harsh UV~field, and considered different limits on the PAH size, above which the C loss is forbidden (limits $N_{\rm C}=0$~(`b'), $N_{\rm C}=60$~(`d'), $N_{\rm C}=\infty$~(`c')), assuming unrestricted H loss. Additionally, we considered the case when no PAH evolution occurs (`a'). We calculated the size, hydrogenation state and charge distributions for PAHs in different positions through the Orion Bar PDR, and utilised the results of the modelling to obtain simulated spectra and band intensities. Our modelling satisfactorily describes the variations of the ratios of $F_{3.6}/F_{11.2}$, $F_{7.7}/F_{11.2}$, $F_{7.7}/F_{3.6}$, $F_{3.3}/F_{3.4}$ with the distance from the ionizing source, even though the amplitudes of these variations are not precisely reproduced. The main conclusions of this work are:

\begin{itemize}
\item The PAH mixture in the Orion Bar PDR consists primarily of large PAHs ($N_{\rm C}\ga80$), with a small fraction of slightly dehydrogenated and normally hydrogenated PAHs ($N_{\rm C}=21$ and 34, respectively) according to our fitting of the ISO spectrum near the DF.  

\item The observed ratios of $F_{7.7}/F_{11.2}$ and $F_{7.7}/F_{3.6}$,  which are believed to characterise the ionization state, have their maximum in the inner part of the PDR bordering the \ion{H}{II} region, and gradually decrease from the PDR towards the MC. The same behaviour is predicted by our model, independent of the model case, while the models `b' and `d' (no C loss and no C loss for PAHs with $N_{\rm C}>60$, respectively) are closer to the observed values. The details of the treatment of the photo-destruction process do not change the ionization state of the PAHs. Our estimates of the ratio $F_{7.7}/F_{11.2}$ are higher than those observed by 2--3 times, apparently due to the low number of neutral PAHs throughout the Orion Bar PDR in our calculations. A more accurate method of computing the PAH ionization potential might eliminate this inconsistency.

\item The observed ratio $F_{3.6}/F_{11.2}$, which is mostly sensitive to the PAH size, has a minimum in the PDR near the IF. In the modelling, we found that the `PDR~2' location (near the DF) is the most adverse place for PAHs, because they can easily become super-hydrogenated, whereas this state is not stable under the harsh UV radiation field. If we assume that all the PAHs can be destroyed, then the depth of the trough of $F_{3.6}/F_{11.2}$ in the PDR is too extreme. However, if we limit the size of PAHs which can be destroyed by $N_{\rm C}=60$ (case `d'), the $F_{3.6}/F_{11.2}$ ratio can be described satisfactorily. 

\item The observed ratio of $F_{3.3}/F_{3.4}$ has a maximum near the IF, and decreases towards the MC by at least a factor of 2. This flux ratio is believed to trace the ratio between aliphatic and aromatic bonds in PAHs. In our model, we adopt H-PAHs as  carriers of the aliphatic bonds. We conclude that the ratio changes as the mean size changes, and it is sensitive to the model case. The behaviour of the ratio is well described if we limit the carbon skeleton destruction by $N_{\rm C}=60$, though our predictions for the `PDR~2' location are lower than those observed, while in the MC they are higher. Despite the inconsistencies, the band at 3.4~$\mu$m can be explained with H-PAHs as carriers.
\end{itemize}

\section*{ACKNOWLEDGEMENTS}
We are very grateful to Gregory Sloan, who provided the UKIRT spectroscopic data, and Olivier Bern\'{e}, who provided the SOFIA photometric data. 

In this work we used the NASA Ames PAH database. Parts of this work are based on observations from the NASA/DLR Stratospheric Observatory for Infrared Astronomy (SOFIA). SOFIA is jointly operated by the Universities Space Research Association, Inc. (USRA), under NASA contract NAS2-97001, and the Deutsches SOFIA Institut (DSI) under DLR contract 50 OK 0901 to the University of Stuttgart. 

Developing of the model of PAHs in UV radiated environment
(Section 3.1), calculations and comparison with observations (Section
4) were supported by the Russian Science Foundation (project
18-13-00269). Construction of a model of expanding PDR around
the Orion HII region in Section 3.2 was supported by the Russian
Science Foundation (project 21-12-00373).

\section*{Data availability}
The model results obtained in this article are available in Figshare at
{\tt\url{https://doi.org/10.6084/m9.figshare.14892612.v1}}.
\bibliographystyle{mnras} 
\bibliography{orion}

\appendix

\section{Absorption cross section of H-PAHs}\label{sect: cabs}
\label{abs_HPAH}

The absorption cross-sections from the work of \citetalias{DL07} are widely used, and generally consistent with mid-IR observations. However, these cross-sections describe only PAHs in the normal hydrogenation state, and they do not include the band at 3.4~$\mu$m, which is needed for interpretation of the observations of the Orion Bar. Some laboratory IR spectra were obtained for tens of PAHs with different numbers of extra hydrogen atoms \citep{sandford13}, but these measurements are not absolute, and we are therefore not able to make full use of them. In the works of \cite{yang13, yang17} the cross-sections for PAHs with aliphatic sidegroups were computed based on the \citetalias{DL07} cross-sections. We follow the method of calculation from their work, but for H-PAHs. Specifically, we modify the cross-sections around the bands at 3.3 and 3.4~$\mu$m, and make them dependent on the PAH hydrogenation level. 

In order to estimate the 3.3 and 3.4~$\mu$m band parameters (central wavelengths, FWHM, and integrated cross-section $\sigma_{\rm band}$) of PAHs in different hydrogenation states we use the NASA Ames IR PAH database. We chose a number of molecules that have IR properties for the normal hydrogenation state, and several states with extra hydrogen atoms: C$_{10}$H$_8$, C$_{10}$H$_8^+$, C$_{16}$H$_{10}$,  C$_{16}$H$_{10}^+$,  C$_{24}$H$_{12}$, C$_{24}$H$_{12}^+$, C$_{54}$H$_{18}^+$, C$_{96}$H$_{24}$. For these molecules, there are different numbers of hydrogenation states in the database, but no less than two for each molecule. We estimate the hydrogenation level using the ratio $X_{\rm H}/X_{\rm H}^{0}$, which indicates the ratio between the number of hydrogen atoms in a specific hydrogenation state and in the normal hydrogenation state of each PAH. This ratio increases from 1 in the initial state to 2 in the most hydrogenated state presented in the database. 

We calculated the cross-sections around the 3.3 and 3.4~$\mu$m bands using the parameters from the database and Lorentzian profiles. The emission bands at 3.3 and 3.4~$\mu$m adopted in the \citetalias{DL07} model are wide, and several thin bands listed in the database could be related to them. We set the bounds from 3.22 to 3.32~$\mu$m and from 3.35 to 3.44~$\mu$m for the bands at 3.3 and 3.4~$\mu$m, respectively, and all thin bands that fall within these bounds are included in the wide bands. We fit all thin bands within the bounds as single bands near 3.3 and 3.4~$\mu$m by Drude profiles, as this profile is adopted in the \citetalias{DL07} model. The exact central positions may vary slightly in different hydrogenation states, but on average the aromatic band is centred at 3.28~$\mu$m, and the aliphatic band at 3.40~$\mu$m. 

\begin{figure}
	\includegraphics[width=0.45\textwidth]{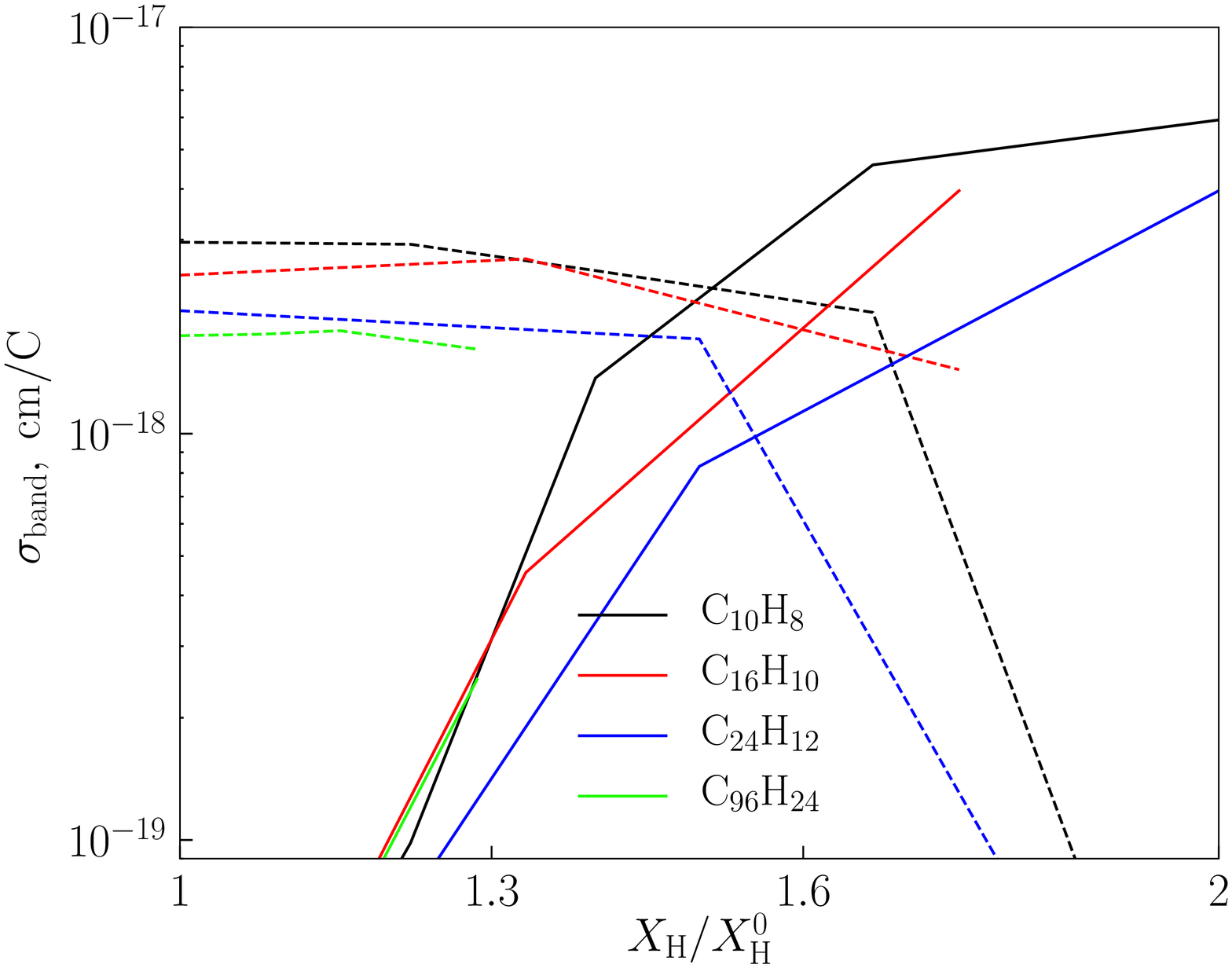}\\
	\includegraphics[width=0.45\textwidth]{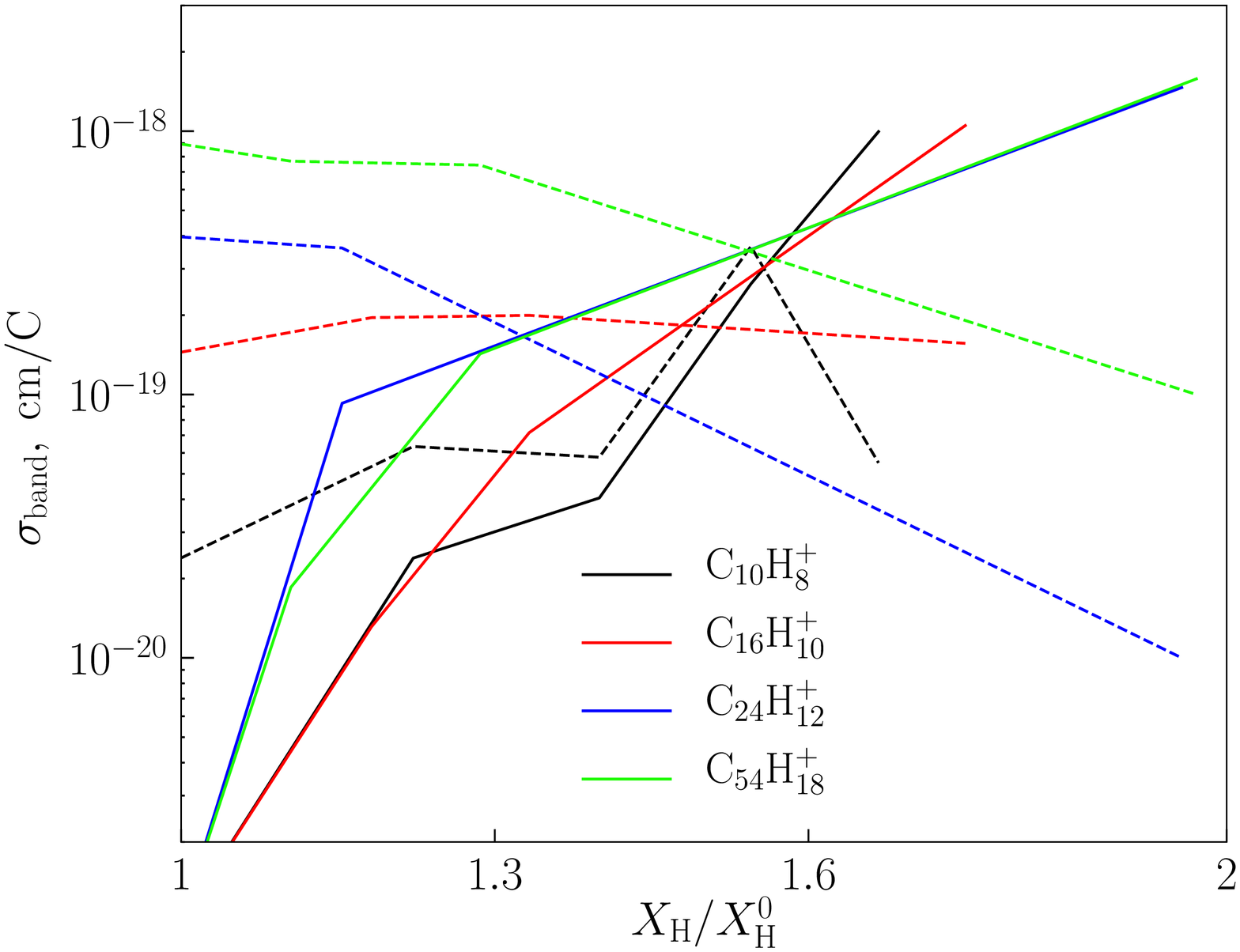}
	\caption{Dependence of integrated cross-sections of the bands at 3.3 and 3.4~$\mu$m on $X_{\rm H}/X_{\rm H}^{0}$ for several PAHs which were chosen from the Nasa Ames PAH IR database. Dashed lines indicate the aromatic (3.3~$\mu$m), solid lines indicate the aliphatic (3.4~$\mu$m) band. On the left, the dependencies for neutral PAHs are shown, on the right --- for cations.}
	\label{fig: cabs}
\end{figure}

For each molecule, we constructed the dependence of the integrated cross-section of the wide bands at 3.3 and 3.4~$\mu$m on $X_{\rm H}/X_{\rm H}^{0}$. These dependencies are shown in Fig.~\ref{fig: cabs}. Due to the significant difference in the IR-properties for ionized and neutral PAHs, we consider these two groups separately. As can be seen in Fig.~\ref{fig: cabs}, the flux of the aromatic band decreases with $X_{\rm H}/X_{\rm H}^{0}$, while the flux of the aliphatic band increases.

The value of $\sigma_{\rm band}$ obtained for a fully aromatic molecule is less than for a fully aliphatic molecule \citep{wexler67, sandford13}, which is related to the relative weakness of the aromatic modes. We fit the dependence for each $i$th molecule presented in Fig.~\ref{fig: cabs} by the function $\sigma_{{\rm band,}i}^{0}(a^{\rm h}_i\cdot X_{\rm H}/X_{\rm H}^{0}+b^{\rm h}_i)$, where $\sigma_{\rm band,i}^{0}$ is the value of the cross-section in the normal hydrogenation state and $a_i$ and $b_i$ are free parameters. We averaged $\sigma_{\rm band,i}^{0}$, $a^{\rm h}_i$, and $b^{\rm h}_i$ for four separate groups: the aromatic band of ionized PAHs, the aromatic band of neutral PAHs, the aliphatic band of ionized PAHs, and the aliphatic band of neutral PAHs, and obtained the mean parameters $\sigma_{\rm band}^{0}$, $a^{\rm h}$, and $b^{\rm h}$ for each group (Table~\ref{tab: abcoef}). We also give the mean FWHM for each case. Finally, $\sigma_{\rm band}^{0}$, $a^{\rm h}$, $b^{\rm h}$ and FWHM were used to calculate the absorption cross-sections of PAHs in different states of ionization and hydrogenation, i.e. we replaced the \citetalias{DL07} cross-sections near 3.3 and 3.4~$\mu$m with bands with the parameters from Table~\ref{tab: abcoef} for each specific state.

\begin{table}
\caption{Mean parameters of the integrated cross-sections and FWHM for the bands at 3.3 and 3.4~$\mu$m for H-PAHs.}
\label{tab: abcoef}
\begin{center}
\begin{tabular}{ccccc}
\hline
Band & $\sigma_{\rm band}^{0}$ & $a^{\rm h}$ & $b^{\rm h}$ & FWHM \\ \hline
3.3, neutral  & $2.23\cdot10^{-18}$ & $-0.76$  & $1.82$ & 0.010 \\
3.3, ionized  & $2.71\cdot10^{-19}$ & $-0.46$  & $1.40$ & 0.010 \\
3.4, neutral  & $3.96\cdot10^{-18}$ & $1.16$   & $-1.29$ & 0.015 \\
3.4, ionized  & $1.26\cdot10^{-18}$ & $2.74$   & $-1.30$ & 0.015 \\
\hline
\end{tabular}
\end{center}
\end{table}

\section{Absorption cross-section of dehydrogenated PAHs}\label{sect: cabs2}

Analogous to H-PAHs, we modify the absorption cross-sections for dehydrogenated PAHs. The NASA Ames IR PAH database includes only a few molecules in the dehydrogenated state, namely C$_{54}$H$_{18}^+$, C$_{66}$H$_{20}$, C$_{66}$H$_{20}^+$, C$_{96}$H$_{24}$, C$_{96}$H$_{24}^+$. Based on the database, we assume that only the bands at 3.3, 11.2 and 12.6~$\mu$m from the list of bands in the \citetalias{DL07} model are primarily sensitive to the decreasing of $X_{\rm H}$. The fitting procedure described in Appendix~\ref{abs_HPAH} was repeated for the dehydrogenated states of the PAHs. As in the previous case, we found the dependence of the integrated cross-section on $X_{\rm H}/X_{\rm H}^{0}$ as $\sigma_{{\rm band,}i}^{0}(a^{\rm d}_i*X_{\rm H}/X_{\rm H}^{0}+b^{\rm d}_i)$, where we adopt $\sigma_{{\rm band,}i}^{0}$ for each molecule equal from \citetalias{DL07}. In the case of dehydrogenated PAHs, the dependencies for ionized and neutral PAHs are similar (only the value of $\sigma_{{\rm band,}i}^{0}$ may differ), so we estimated the coefficients $a^{\rm d}$ and $b^{\rm d}$ averaged for all molecules, regardless their charge (Table~\ref{tab: abcoef_dehydr}). In the \citetalias{DL07} model, each of the bands at 11.2 and 12.6~$\mu$m consists of two thin bands, so we modify $\sigma_{\rm band}$ of these thin bands similarly, without changing their FWHM. 

\begin{table}
\caption{Mean parameters of the integrated cross-sections and FWHM for the bands and 3.3, 11.2, and 12.6~$\mu$m bands for dehydrogenated PAHs.}
\label{tab: abcoef_dehydr}
\begin{center}
\begin{tabular}{cccc}
\hline
Band & $a^{\rm d}$ & $b^{\rm d}$ & FWHM \\ \hline
3.3  & 0.948  & 0.003 & 0.008 \\
11.2 & 0.987  & 0.041 & - \\
(\citetalias{DL07}: 11.23, 11.33) & & \\
12.6 & 0.799   & 0.200 & - \\
(\citetalias{DL07}: 12.62, 12.69) & & \\
\hline
\end{tabular}
\end{center}
\end{table}

\bsp	
\label{lastpage}
\end{document}